\documentclass[lettersize,journal]{IEEEtran}
\usepackage{amsmath,amsfonts}
\usepackage{algorithmic}
\usepackage{algorithm}
\usepackage{array}
\usepackage[caption=false,font=normalsize,labelfont=sf,textfont=sf]{subfig}
\usepackage{textcomp}
\usepackage{stfloats}
\usepackage{url}
\usepackage{verbatim}
\usepackage{graphicx}
\usepackage{booktabs}
\usepackage{cite}
\usepackage{multirow,multicol}
\usepackage{xcolor}

\begin{document}

\title{A Differentiable Framework for Full and Phaseless Data Inversion Using Neural Implicit Contrast-Source Representation}

\author{Haoran Sun, \IEEEmembership{Graduate Student Member}, Daoqi Liu, \IEEEmembership{Graduate Student Member}, Hongyu Zhou, \IEEEmembership{Graduate Student Member}, Maokun Li, \IEEEmembership{Fellow, IEEE}, Shenheng Xu, \IEEEmembership{Member, IEEE}, and Fan Yang, \IEEEmembership{Fellow, IEEE}
\thanks{This work was supported by Noncommunicable Chronic Diseases-National Science and Technology Major Project under Grant 2024ZD0530000.}
\thanks{Haoran Sun, Daoqi Liu, Hongyu Zhou, Maokun Li, Shenheng Xu and Fan Yang are with the Department of Electronic Engineering, Beijing National Research Center for Information Science and Technology (BNRist), and State Key laboratory of Space Network and Communications, Tsinghua University, Beijing 100084, China. (e-mail: maokunli@tsinghua.edu.cn).}}

\markboth{Journal of \LaTeX\ Class Files,~Vol.~14, No.~8, August~2021}%
{Shell \MakeLowercase{\textit{et al.}}: A Sample Article Using IEEEtran.cls for IEEE Journals}

\IEEEpubid{0000--0000/00\$00.00~\copyright~2021 IEEE}

\maketitle
\begin{abstract}
      In this study, we extend the contrast source inversion to a fully differentiable, unsupervised framework based on a neural implicit representation of the contrast source. Specifically, instead of a pixel-wise discrete representation, the contrast source is parameterized by a lightweight residual multilayer perceptron (ResMLP) as a continuous neural field conditioned on spatial coordinates and transmitter settings. This continuous parameterization provides a more flexible representation of the contrast source and improves reconstruction accuracy and robustness under noisy measurements. Building on this representation, the state equation and data equation are combined with total-variation regularization to form a differentiable objective function. By reformulating the VIE-constrained inversion as an end-to-end differentiable optimization problem, the network parameters and the medium contrast are jointly optimized via automatic differentiation. Within the same framework, both full and phaseless data inversion are accommodated by only modifying the data misfit function. Numerical experiments demonstrate that this scheme yields higher reconstruction accuracy and robustness than conventional CSI across a range of noise levels and measurement settings. The continuous neural field further enables super-resolution inference at resolutions finer than the training grid, decoupling inversion cost from reconstruction fidelity. Ablation studies and comparisons with alternative neural architectures further confirm that the contrast source parameterization and VIE-based formulation are both essential to the observed improvements.
\end{abstract}


\section{Introduction}
\IEEEPARstart{E}{lectromagnetic} inverse scattering aims to reconstruct material properties within a domain of interest (DOI) from measured field data. It has been widely applied to electromagnetic (EM) imaging \cite{abubakar2003two,xu2022learning}, nondestructive testing \cite{yang2024layered}, geophysical exploration \cite{abubakar20082}, medical diagnostics \cite{bevacqua2021millimeter}, etc. In practice, however, measurements are corrupted by noise and may lack or be unreliable in phase information. Coupled with intrinsic nonlinearity and ill-posedness, EM inverse scattering problems pose significant challenges for accurate and robust reconstruction.

EM inversion methods, including classical numerical solvers and their learning-augmented variants, are typically built upon grid-based discretizations of both unknown fields and material distributions\cite{chen2018computational,liu2022towards,salucci2022artificial}. Under this paradigm, these methods can be broadly grouped into deterministic, stochastic, and learning-enhanced methods. Deterministic methods typically rely on electromagnetic scattering physics and enforce consistency between measured and simulated data through linear or nonlinear programming \cite{devaney1982inversion,habashy1993beyond,colton2003linear,chew1990reconstruction,cui2001inverse,van2001contrast,chen2009subspace}. Stochastic inversion methods reconstruct unknown parameters stochastically exploring the posterior distribution \cite{yin2022interior,huang2023full}. It can better avoid local minima and quantify uncertainty, but the computational efficiency is usually lower than that of deterministic methods. Recently, Deep learning (DL) has been incorporated to enhance or accelerate inversion \cite{liu2022towards}, usually through fully data-driven mappings and deep unrolling architectures. Fully data-driven methods learn a direct mapping from measurements to material parameters and can serve as fast surrogate solvers within the training distribution \cite{wei2018deep,li2018deepnis,sun2022joint,sun2025PhysicsInformed}. Deep unrolling networks unfold iterative inversion procedures and embed the forward operator to improve interpretability and generalization \cite{guo2021physicsisp,deshmukh2022unrolled,shan2022neural,zhang2022unrolled,liu2022som,liu2025multi}. Overall, these methods usually rely on discrete spatial parameterizations for both fields and material properties. Whether using pixel-wise discrete basis functions in iterative schemes or feature maps in neural networks (NNs), unknown fields are represented discretely. This creates a representation gap, in which continuous, typically smooth physical fields are only approximated by finite-resolution sampling. Under sparse or noisy measurements, grid-induced artifacts can be amplified, lowering reconstruction accuracy and robustness.

\IEEEpubidadjcol

An alternative is to represent unknown fields using neural implicit functions, as explored in physics-informed neural networks (PINNs) and related unsupervised schemes \cite{bar2019unsupervised,hu2023more,hu2024priori,wang2024universal,wang2025deep}. These methods parameterize the unknown field using neural networks and incorporate governing equations into the loss function, without requiring paired training data. Related advances have also been reported in seismic full-waveform inversion, where physics-informed neural networks and coordinate-based neural representations have been used for wavefield reconstruction, implicit model parameterization, and source-independent inversion \cite{Song2022WRI,Sun2023ImplicitFWI,Song2026PCDLFWI}, further demonstrating the potential of neural implicit and physics-informed representations in wave-based inverse problems. However, most PINNs-based inverse scattering approaches enforce physics through residuals of partial differential equations (PDEs) and boundary conditions (BCs) \cite{hu2023more,hu2024priori}. In these methods, the optimization process is often sensitive to sampling strategies, network capability, and the balancing among multiple loss terms \cite{lau2024pinnacle}. These issues become more pronounced in strongly nonlinear and ill-posed problems or under noisy measurements, frequently leading to slow convergence and extensive hyperparameter tuning.

In this work, we propose a physics-informed deep contrast source inversion framework (DeepCSI) inspired by the classical contrast source inversion formulation \cite{van1997contrast}. DeepCSI performs unsupervised, case-by-case inversion, where the network optimization itself constitutes the inversion process. Specifically, the contrast sources are parameterized by a lightweight ResMLP as a continuous neural implicit field conditioned on spatial coordinates and transmitter locations, thereby enabling a more faithful representation of smooth contrast source distributions. This neural representation also provides implicit regularization that suppresses noise-induced nonphysical artifacts. The medium contrast is modeled as learnable tensors and is jointly optimized with the network parameters. The hybrid loss function combines the state equation, data equation, and total-variation regularization. The resulting fully differentiable framework supports automatic-differentiation-based updates without manual gradient derivations, and accommodates both full and phaseless measurements by modifying only the data-consistency term. In this manner, DeepCSI is constrained by the volume integral equation (VIE), where the Green's function operator explicitly characterizes wave propagation. This avoids enforcing physics via PDEs and boundary conditions with many competing loss terms, thereby accelerating convergence. Numerical experiments demonstrate that DeepCSI consistently outperforms conventional CSI in both accuracy and robustness across different noise levels and measurement settings. The main contributions are summarized as follows:

\begin{enumerate}
    \item A neural implicit contrast source representation is incorporated into CSI, introducing implicit regularization that improves modeling fidelity and robustness to noise-induced artifacts while enabling super-resolution inference beyond the training grid resolution.
    \item A fully differentiable inversion framework is developed by integrating the VIE operators into an end-to-end differentiable computational graph, enabling joint optimization via automatic differentiation and seamless extension from full to phaseless data by modifying only the data-consistency term.
    \item Extensive experiments, including ablation studies and comparisons with both conventional methods and learning-based alternatives, demonstrate the superiority of DeepCSI and confirm that the neural representation is the primary factor in artifact suppression.
\end{enumerate}

This paper is organized as follows. Section II briefly reviews CSI. Section III discusses DeepCSI, including both the ResMLP-based contrast source representation and the differentiable inversion formulation for full and phaseless data. Section IV reports numerical results. Section V concludes the paper.

\section{Forward Modeling and Contrast Source Inversion}
\begin{figure}[t]
    \centering
    \includegraphics[width=0.35\textwidth]{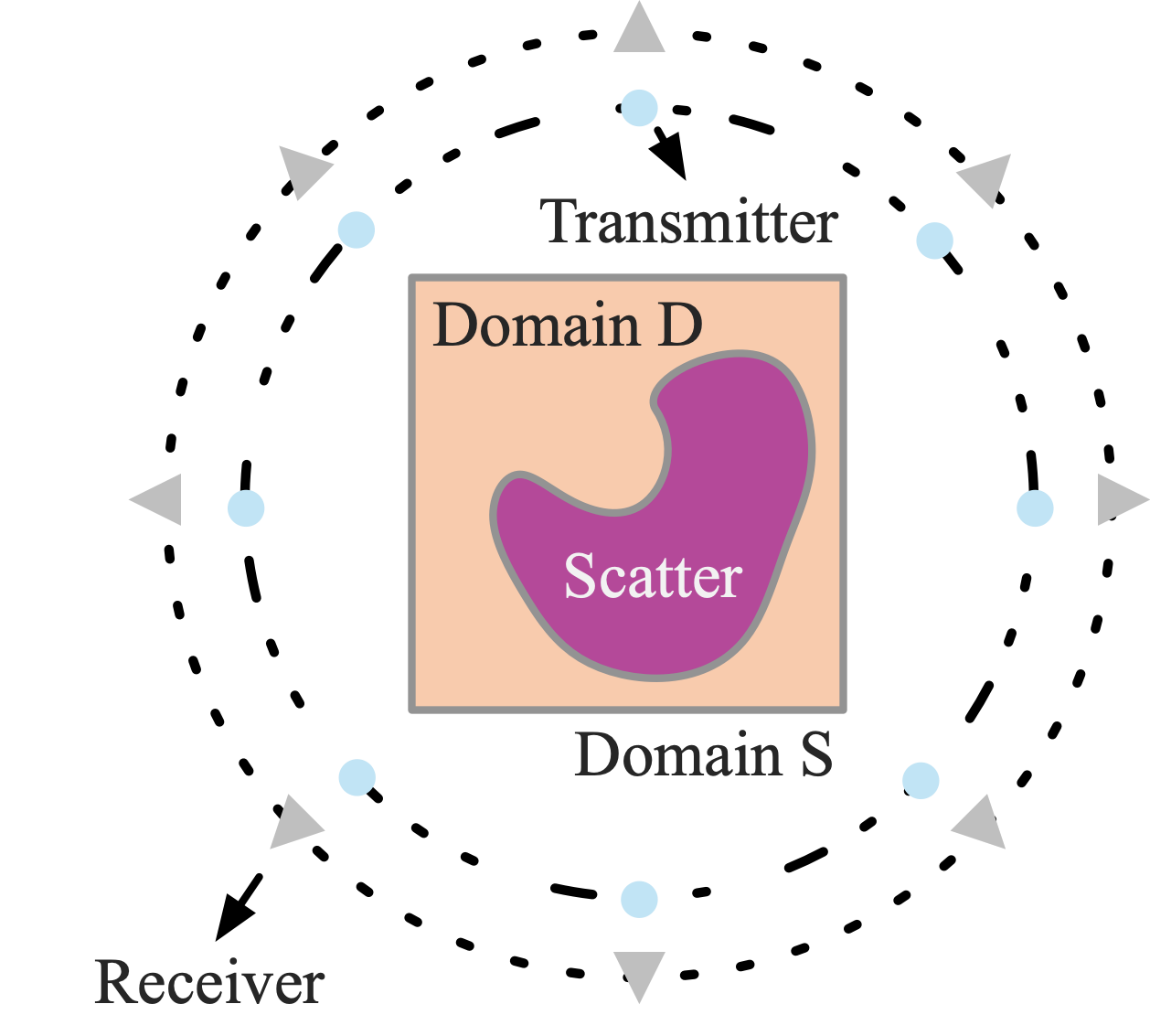}
    \caption{Typical setup of 2-D inverse scattering problems.}
    \label{figs_setup}
\end{figure}

\begin{figure*}[t]
    \centering
    \includegraphics[width=0.85\textwidth]{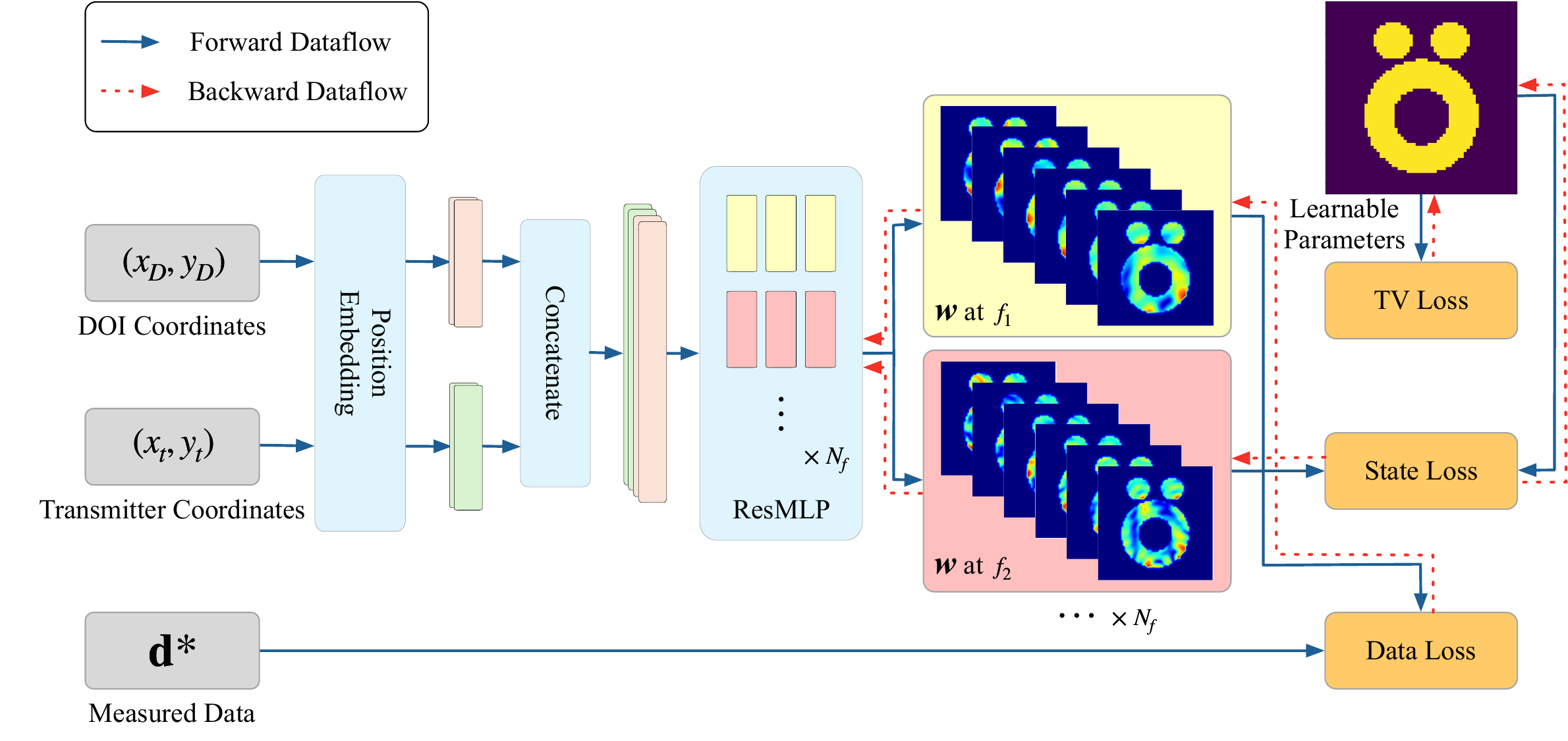}
    \caption{The framework of DeepCSI. Each subnet in the ResMLP module predicts the contrast source distribution at a specific frequency, conditioned on the DOI coordinates~$\rho$ and the transmitter coordinates~$\rho_t$. The subnets share the same architecture but have independent weights. $\mathbf{d}^*$ denotes the measured scattered field for FD inversion and the measured total-field magnitude for PD inversion.}
    \label{figs_framework}
\end{figure*}
\subsection{Volume Integral Equation Formulation}
Consider a two-dimensional (2D) DOI in free space, where unknown scatterers are illuminated by TM-polarized line sources arranged in a circular configuration. The total field is measured by receivers placed around the DOI, as illustrated in Fig. \ref{figs_setup}. Under this setup, the incident field $E^{inc}$, scattered field $E^{sca}$, and total field $E^{tot}$ satisfy the following equations \cite{jin2015theory}:
\begin{equation}
	\begin{aligned}
		& E^{t o t}(\boldsymbol{\rho}) \\
		& =E^{i n c}(\boldsymbol{\rho})+k_b^2 \int_D G_D\left(\boldsymbol{\rho}, \boldsymbol{\rho}^{\prime}\right) \chi\left(\boldsymbol{\rho}^{\prime}\right) E^{t o t}\left(\boldsymbol{\rho}^{\prime}\right) d \boldsymbol{\rho}^{\prime}, \boldsymbol{\rho} \in D
	\end{aligned}
\end{equation}
\begin{equation}
E^{s c a}(\boldsymbol{\rho})=k_b^2 \int_D G_S\left(\boldsymbol{\rho}, \boldsymbol{\rho}^{\prime}\right) \chi\left(\boldsymbol{\rho}^{\prime}\right) E^{t o t}\left(\boldsymbol{\rho}^{\prime}\right) d \boldsymbol{\rho}^{\prime}, \boldsymbol{\rho} \in S
\end{equation}
where $\boldsymbol{\rho}=(x,y)$ denotes the position in 2D space, $k_b=\omega\sqrt{\mu_{0}\epsilon_0}$ is the wavenumber in free space, $\chi=\epsilon_r/\epsilon_0 -1$ represents the scatterer contrast, and $\epsilon_r$ is the relative permittivity. $G_D$ and $G_S$ denote the Green's functions of the background, while $D$ and $S$ correspond to the DOI and the observation domain respectively. By discretizing the DOI into $N\times N$ subdomains and applying pulse basis functions along with a point-matching strategy, the integral equation can be transformed into a linear system of equations, which can be expressed as \cite{harrington1993field}:
\begin{equation}
	\mathbf{E}^{t o t}=\mathbf{E}^{i n c}+\mathbf{G}_D \boldsymbol{\chi} \mathbf{E}^{t o t}
	\label{eq:state}
\end{equation}
\begin{equation}
	\mathbf{E}^{s c a}=\mathbf{G}_S \boldsymbol{\chi} \mathbf{E}^{t o t}
	\label{eq:data}
\end{equation}
where $\mathbf{E}^{{tot}}$, $\mathbf{E}^{{inc}}$, and $\mathbf{E}^{{sca}}$ denote the total, incident, and scattered field vectors, respectively. Eqs. \eqref{eq:state} and \eqref{eq:data} describe the electromagnetic field interactions within and outside the DOI, forming the physical foundation for inverse scattering analysis.

ISPs aim to reconstruct the unknown contrast distribution $\boldsymbol{\chi}$ from the measured scattered field $\mathbf{d}^*$. However, since the total field $\mathbf{E}^{{tot}}$ depends nonlinearly on $\boldsymbol{\chi}$ through the integral equation, the ISP is inherently nonlinear and ill-posed. Small measurement perturbations may lead to large variations in the reconstructed contrast, resulting in instability and slow convergence in conventional iterative methods. To overcome these challenges, the contrast source inversion method reformulates the ISP into a coupled, physically consistent optimization framework, as discussed in the following subsection.

\subsection{Contrast Source Inversion}
To mitigate the nonlinearity and ill-posedness of the inverse scattering problem, the CSI introduces an auxiliary variable known as the \emph{contrast source}, which is defined as:
\begin{equation}
    \label{eq:omega_chi}
	\boldsymbol{\omega} = \boldsymbol{\chi} \mathbf{E}^{t o t}
\end{equation}
which represents the equivalent current induced within the scatterer. Substituting this definition into the integral formulation leads to two coupled equations:
\begin{equation}
	\boldsymbol{\omega}=\boldsymbol{\chi}\mathbf{E}^{i n c}+\boldsymbol{\chi}\mathbf{G}_D \boldsymbol{\omega}
	\label{eq:state_simp}
\end{equation}
\begin{equation}
	\mathbf{E}^{s c a}=\mathbf{G}_S \boldsymbol{\omega}
	\label{eq:data_simp}
\end{equation}
where Eq. \eqref{eq:state_simp} is typically referred to as the state equation, Eq. \eqref{eq:data_simp} is the data equation \cite{van2001contrast}. CSI reconstructs both the contrast $\boldsymbol{\chi}$ and the contrast source $\boldsymbol{\omega}$ by minimizing the misfits between the state and data equations. A typical objective function can be expressed as \cite{van1997contrast}:
\begin{equation}
    \begin{aligned}
        \mathcal{L}_{C S I}(\boldsymbol{\omega}, \boldsymbol{\chi})&=\mathcal{L}_D(\boldsymbol{\omega}, \boldsymbol{\chi})+\mathcal{L}_S(\boldsymbol{\omega}) \\ 
        &=\frac{\left\|\boldsymbol{\omega}-\boldsymbol{\chi} \mathbf{E}^{i n c}-\boldsymbol{\chi} \mathbf{G}_D \boldsymbol{\omega}\right\|_2^2}{\left\|\boldsymbol{\chi} \mathbf{E}^{i n c}\right\|_2^2}+\frac{\left\|\mathbf{E}^{s c a}-\mathbf{G}_S \boldsymbol{\omega}\right\|_2^2}{\left\|\mathbf{E}^{s c a}\right\|_2^2}
    \end{aligned}
\end{equation}

Additionally, multiplicative regularization terms such as the weighted L2 total variation (TV) regularization have been introduced to improve the inversion accuracy \cite{van2001contrast}, whose expression is:
\begin{equation}
L_{T V}(\boldsymbol{\chi})=\frac{\sum_{i=1}^{N^2}\left(\left\|\nabla \boldsymbol{\chi}_i\right\|^2+\delta^2\left(\boldsymbol{\chi}^{\prime}\right)\right)^{\frac{p}{2}}}{\sum_{i=1}^{N^2}\left(\left\|\nabla \boldsymbol{\chi}_i^{\prime}\right\|^2+\delta^2\left(\boldsymbol{\chi}^{\prime}\right)\right)^{\frac{p}{2}}}
\end{equation}
\begin{equation}
    \delta^2(\boldsymbol{\chi}^{\prime})=\mathcal{L}_D(\boldsymbol{\omega}, \boldsymbol{\chi}^{\prime})\cdot\Delta^2
\end{equation}
where $p$ denotes the norm of TV regularization, which is set to $2$ in this paper. $\Delta$ represents the grid size of the discrete domain $D$. $\boldsymbol{\chi}$ and $\boldsymbol{\chi}^{\prime}$ denote the contrast at current and previous iterations, respectively. The cost function of the multiplicative regularization contrast source inversion (MRCSI) can be expressed as \cite{abubakar2005application}:
\begin{equation}
    \mathcal{L}_{MRCSI}(\boldsymbol{\omega}, \boldsymbol{\chi}) = \mathcal{L}_{CSI}(\boldsymbol{\omega}, \boldsymbol{\chi})L_{T V}(\boldsymbol{\chi})
\end{equation}

In solving this optimization problem, CSI minimizes the cost function by alternately updating the contrast source $\boldsymbol{\omega}$ and the contrast $\boldsymbol{\chi}$. Through this iterative process, CSI achieves quantitative reconstruction of the permittivity distribution while maintaining physical consistency between the electromagnetic fields and measured data. The update expression for the contrast source at the $n$-th iteration is given by:
\begin{equation}
    \boldsymbol{\omega}_n = \boldsymbol{\omega}_{n-1}+\alpha_n^{\boldsymbol{\omega}}\mathcal{V}_n
\end{equation}
where $\mathcal{V}_n$ is the Polak–Ribi\`ere conjugate gradient direction, which is the gradient of the cost function with respect to the contrast source. The step size $\alpha_n^{\boldsymbol{\omega}}$ for updating the contrast source is obtained via a line search method. After updating the contrast source, the total field can be updated as:
\begin{equation}
    \mathbf{E}^{tot}_n=\mathbf{E}^{inc}+\mathbf{G}_D\boldsymbol{\omega}_n
\end{equation}
Subsequently, the contrast can be updated as:
\begin{equation}
    \boldsymbol{\chi}_n = \boldsymbol{\chi}_{n-1}+\alpha_n^{\boldsymbol{\chi}} \mathcal{D}_n
\end{equation}
where the update direction $\mathcal{D}_n$ is the Polak–Ribi\`ere conjugate gradient direction, representing the gradient of the cost function with respect to the contrast. The step size $\alpha_n^{\boldsymbol{\chi}}$ for contrast update is similarly determined via a line search method. Details can be found in \cite{van2003multiplicative}.

Although CSI provides a physically consistent and quantitatively accurate framework, it models the contrast source distribution on a discrete grid and lacks continuity and smoothness constraints on contrast sources, which can degrade inversion accuracy in complex scenarios. To overcome these limitations, a fully differentiable inversion framework based on neural implicit contrast source representation, DeepCSI, is introduced in the next section.

\section{Physics-Informed Deep Contrast Source Inversion Framework}
In this section, we propose DeepCSI, a physics-informed deep contrast source inversion framework that extends the conventional CSI approach into a fully differentiable optimization paradigm. As illustrated in Fig. \ref{figs_framework}, DeepCSI employs a ResMLP as a global basis function to represent the contrast source distribution, mapping spatial coordinates $\boldsymbol{\rho}$ to the contrast source $\boldsymbol{\omega}(\boldsymbol{\rho})$ and enabling end-to-end differentiable computation. By modeling the target properties as learnable tensors and introducing a hybrid loss function that integrates the state equation, data equation, and total variation regularization, DeepCSI reformulates the ISP into a differentiable optimization problem that jointly enforces physical consistency and data fidelity. Leveraging automatic differentiation and stochastic optimizers in modern deep learning frameworks \cite{paszke2019pytorch}, DeepCSI simultaneously updates the network parameters and medium properties, achieving accurate, stable, and flexible inversion reconstruction under various measurement conditions.

\begin{algorithm}[t]
\caption{DeepCSI for Full and Phaseless Data Inversion}
\label{alg:deepcsi}
\textbf{Input:} Measured data $\mathbf{d}^{\ast}$, incident field $\mathbf{E}^{inc}$, measurement type $\mathrm{mode}\in\{\mathrm{FD},\mathrm{PD}\}$, Green's function $\mathbf{G}_D$ and $\mathbf{G}_S$

\textbf{Require:} Frequency set $\mathcal{F}$, transmitter locations $\{\boldsymbol{\rho}_t\}$, DOI coordinates $\{\boldsymbol{\rho}\}$, Positional encoding bandwidth $M$, iteration number $N_{\mathrm{iter}}$, TV weight schedule $\alpha(n)$, Adam settings for network parameters $\theta$ and medium contrast $\boldsymbol{\chi}$.

\textbf{Output:} reconstructed contrast $\boldsymbol{\chi}$.

\begin{algorithmic}[1]
\STATE \textbf{do}
\STATE Initialize $\boldsymbol{\chi}^{(0)}=\mathbf{0}$ and ResMLP subnet parameters $\{\theta_f\}_{f\in\mathcal{F}}$, set Adam optimizers for $\theta$ and $\boldsymbol{\chi}$ as Adam$_\theta$ and Adam$_{\boldsymbol{\chi}}$.
\STATE \textbf{done}
\FOR{$n=0,1,\ldots,N_{\mathrm{iter}}-1$}
    \STATE Set $\alpha \leftarrow \alpha(n)$.
    \FOR{each frequency $f\in\mathcal{F}$}
        \STATE Compute / input $u \leftarrow \left[\gamma(\boldsymbol{\rho}),\gamma(\boldsymbol{\rho}_t)\right]$ on all grid points $\boldsymbol{\rho}\in D$.
        \STATE Predict contrast source $\boldsymbol{\omega}(\boldsymbol{\rho}; \boldsymbol{\rho}_t, f) \leftarrow \boldsymbol{\Phi}_{\theta_f}(u)$.
    \ENDFOR
    \STATE Stack $\boldsymbol{\omega} \leftarrow \{\boldsymbol{\omega}(\boldsymbol{\rho}; \boldsymbol{\rho}_t, f)\}$ over all transmitters and frequencies.

    \STATE Compute state residual
    \STATE \hspace{1em} $\mathbf{r}_S \leftarrow \boldsymbol{\omega}-\boldsymbol{\chi}^{(n)}\mathbf{E}^{inc}-\boldsymbol{\chi}^{(n)}\mathbf{G}_D\boldsymbol{\omega}$.

    \IF{$\mathrm{mode}=\mathrm{FD}$}
        \STATE Compute data residual
        \STATE \hspace{1em} $\mathbf{r}_D \leftarrow \mathbf{d}^{\ast}-\mathbf{G}_S\boldsymbol{\omega}$.
    \ELSIF{{$\mathrm{mode}=\mathrm{PD}$}}
        \STATE Compute data residual for phaseless measurements
        \STATE \hspace{1em} $\mathbf{r}_D \leftarrow \mathbf{d}^{\ast}-\left|\mathbf{E}^{inc}_r+\mathbf{G}_S\boldsymbol{\omega}\right|$.
    \ENDIF
    \STATE Compute loss
        \STATE \hspace{1em} $\mathcal{L}\leftarrow
        \dfrac{\|\mathbf{r}_S\|_2}{\|\boldsymbol{\chi}^{(n)}\mathbf{E}^{inc}\|_2}
        +\dfrac{\|\mathbf{r}_D\|_2}{\|\mathbf{d}^{\ast}\|_2}
        +\alpha\sum|\nabla\boldsymbol{\chi}^{(n)}|$.

    \STATE Backpropagate gradients by automatic differentiation.
    \STATE Update $\{\theta_f\}$ using Adam$_\theta$, update $\boldsymbol{\chi}^{(n)}$ using Adam$_\chi$, obtain $\boldsymbol{\chi}^{(n+1)}$.
\ENDFOR

\STATE \textbf{return} $\boldsymbol{\chi}^{(N_{\mathrm{iter}})}$.
\end{algorithmic}
\end{algorithm}


\subsection{ResMLP-based Contrast Source Representation}
\begin{figure}[t]
    \centering
    \includegraphics[width=0.45\textwidth]{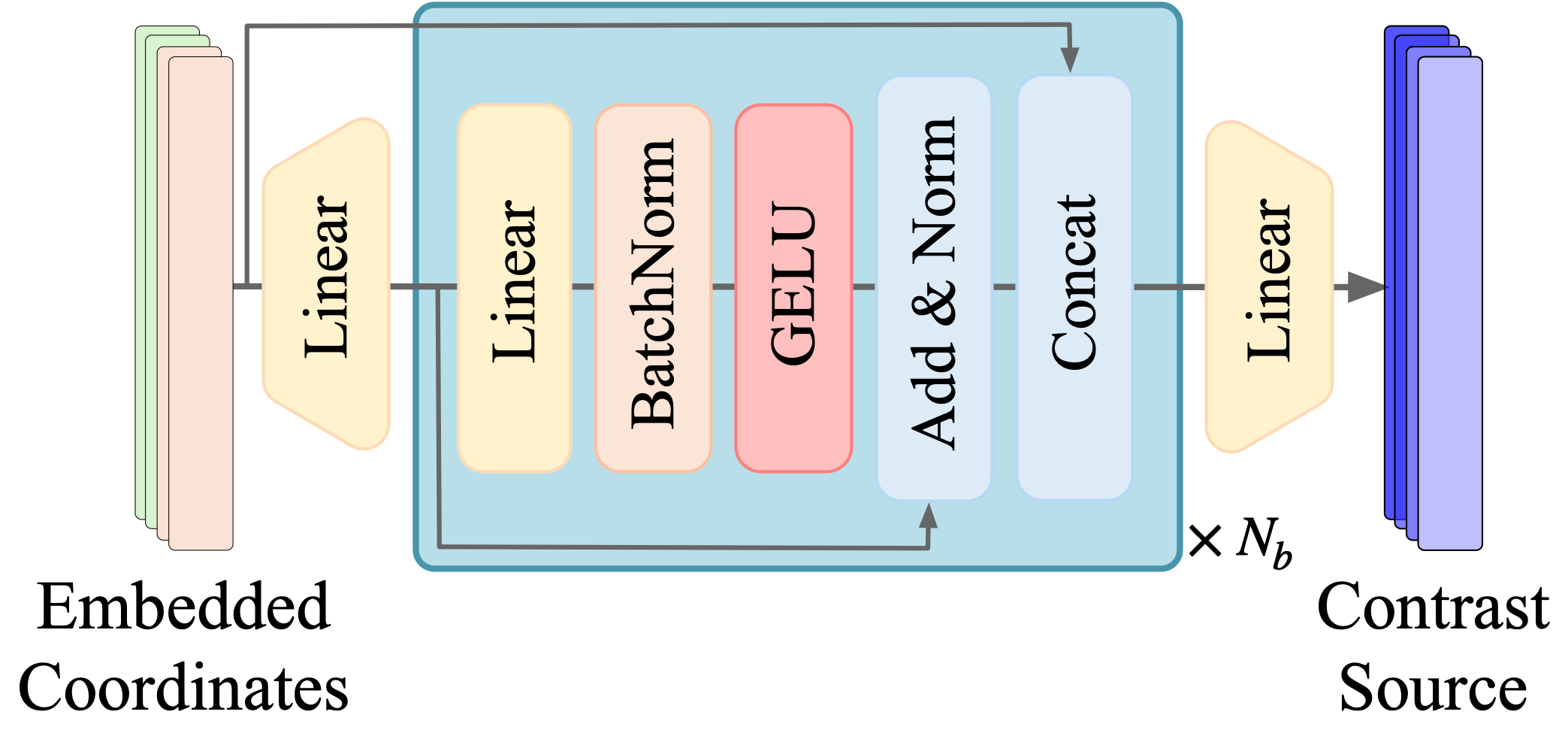}
    \caption{The architecture of the ResMLP in DeepCSI.}
    \label{figs_resmlp}
\end{figure}

In the DeepCSI framework, the ResMLP is used to represent the contrast source distribution in DOI, as illustrated in Fig. \ref{figs_framework}, thereby leveraging the continuity and smoothness of neural implicit representations to model the contrast source more faithfully. Unlike pixel-wise discrete representations in which each spatial location stores an independent, directly optimizable variable, the neural implicit representation defines a continuous mapping from spatial coordinates to contrast source values through shared network parameters~$\theta$. Because all spatial locations share the same parameters, updating~$\theta$ at any training point simultaneously affects predictions at all other locations, providing an implicit smoothness prior that suppresses nonphysical high-frequency artifacts. The network can also be evaluated at arbitrary non-grid coordinates, enabling inference at resolutions finer than the training grid. For each transmitter, the ResMLP takes the concatenation of the spatial coordinates in the DOI and the coordinates of the current transmitter as input and predicts the corresponding contrast source. Since the contrast sources vary across frequencies, multiple independent ResMLP subnets with identical architectures but separate parameters are adopted for different frequencies, ensuring accurate modeling of the contrast source. Meanwhile, to address the spectral bias inherent in standard MLPs \cite{rahaman2019spectral}, a frequency-domain positional encoding is applied to the spatial coordinates as \cite{mildenhall2021nerf}:
\begin{equation}
\gamma(x, y)=\left[\begin{array}{l}
\cos (2 x), \sin (2 x), \ldots, \cos \left(2^M x\right), \sin \left(2^M x\right) \\
\cos (2 y), \sin (2 y), \ldots, \cos \left(2^M y\right), \sin \left(2^M y\right)
\end{array}\right]^T
\end{equation}
where $M$ is a hyperparameter that controls the spectral bandwidth of the input coordinate vector. After positional encoding, the concatenated input $[\gamma(\boldsymbol{\rho}), \gamma(\boldsymbol{\rho}_t)]$ is fed into the ResMLP to capture high-frequency variations of the contrast source.

The architecture of each ResMLP subnet is illustrated in Fig. \ref{figs_resmlp}. Each subnet comprises $N_b$ stacked residual modules, and each module contains a linear layer with 256 hidden units, batch normalization, a GELU activation, and residual connections. The network outputs two channels representing the real and imaginary parts of the contrast source. After reshaping and combining these channels, the contrast source distribution $\boldsymbol{\omega}$ is obtained, as shown in Fig. \ref{figs_framework}.

In the multi-frequency setting, cross-frequency physical consistency is enforced through the shared contrast tensor~$\chi$, which is jointly optimized across all frequencies. Since the contrast source $\omega_f = \chi \cdot E^{\mathrm{tot}}_f$ varies across frequencies due to the frequency-dependent total field rather than changes in the scatterer, independent subnets are a natural design choice that matches this physical structure. An alternative would be a shared-weight conditional architecture in which frequency~$f$ serves as an additional network input. While this design could reduce the parameter count, it risks spectral interference between frequencies with distinct spatial patterns, capacity bottlenecks in a fixed-width network, and gradient conflicts during backpropagation. In practice, each ResMLP subnet contains approximately 540K parameters, occupying only about 2.1~MB in float32, whereas the Green's function matrix~$\mathbf{G}_D$ alone requires approximately 129~MB for a $64 \times 64$ grid. Furthermore, the physical DOI coordinates are used without frequency-dependent rescaling, as each subnet implicitly learns the spatial scale appropriate to its operating frequency through the frequency-dependent Green's functions in the loss function.


\subsection{Differentiable Contrast Inversion}

In DeepCSI, the ISP is formulated as a differentiable optimization problem over the contrast source $\boldsymbol{\omega}$ and the contrast $\boldsymbol{\chi}$. By treating the contrast $\boldsymbol{\chi}$ as a learnable tensor and parameterizing the contrast source $\boldsymbol{\omega}$  with a ResMLP-based neural implicit field, the entire inversion can be optimized end-to-end within modern deep-learning frameworks.

During inversion, the parameters of all ResMLP subnets are optimized jointly with the medium properties, so that the training of ResMLP directly corresponds to the inversion process. Specifically, the VIE-constrained inverse scattering problem is formulated as a single differentiable computational graph. Automatic differentiation, as provided by PyTorch~\cite{paszke2019pytorch}, serves as the computational engine that propagates gradients end-to-end through this graph, replacing the analytical gradient derivations and manual update rules required in traditional CSI.

In all experiments, two Adam optimizers \cite{kingma2014adam} are employed to update the ResMLP parameters and the medium properties, with initial learning rates set to 0.05 and 0.1, respectively. The learning rates gradually decay following an exponential schedule, and at $n$-th iteration they are scaled by $0.1^{n/2000}$. This differentiable inversion strategy enables DeepCSI to flexibly handle various measurement configurations while preserving physical consistency between the reconstructed contrast sources and the measured data.


\subsection{Loss Function for Full and Phaseless Data}
The loss function used in DeepCSI follows the principle of CSI but is implemented in a differentiable form, enabling a unified framework for both full-data (FD) and phaseless-data (PD) inversion. When the measurement data $\mathbf{d}^*$ contain both amplitude and phase information, referred to as the full-data, the loss function is expressed as:
\begin{equation}
\begin{aligned}
    \mathcal{L}^{f d}&=\mathcal{L}_S^{f d}+\mathcal{L}_D^{f d}+\alpha \mathcal{L}_{T V}^{f d} \\
    &=\frac{\left\|\boldsymbol{\Phi}_\theta(\gamma(\boldsymbol{\rho}))-\boldsymbol{\chi} \mathbf{E}^{i n c}-\boldsymbol{\chi} \mathbf{G}_D \boldsymbol{\Phi}_\theta(\gamma(\boldsymbol{\rho})) \right\|_2}{\left\|\boldsymbol{\chi} E^{i n c}\right\|_2} \\
    &+\frac{\left\|\mathbf{d}^*-\mathbf{G}_S \boldsymbol{\Phi}_\theta(\gamma(\boldsymbol{\rho}))\right\|_2}{\left\|\mathbf{d}^*\right\|_2}+\alpha \sum|\nabla \boldsymbol{\chi}|
\end{aligned}
\end{equation}
where $\boldsymbol{\Phi}_\theta(\gamma(\boldsymbol{\rho}))$ represents the contrast source distribution predicted by the ResMLP, $\theta$ are the learnable parameters, and $\alpha$ controls the weight of the total variation (TV) regularization term. In this study, $\alpha$ is initialized to 0.1 and linearly decays to 0.05 during inversion to balance data fidelity and sharp boundaries. 


In practical EM sensing, accurate phase measurements are often challenging, especially at high frequencies. To accommodate such scenarios, DeepCSI can be directly adapted to phaseless-data inversion by modifying only the data equation term in the loss function, without any additional gradient derivation or code modification. The corresponding loss function is given by:
\begin{equation}
\begin{aligned}
    \mathcal{L}^{p d}&=\mathcal{L}_S^{p d}+\mathcal{L}_D^{p d}+\alpha \mathcal{L}_{T V}^{p d} \\
    &=\frac{\left\|\boldsymbol{\Phi}_\theta(\gamma(\boldsymbol{\rho}))-\boldsymbol{\chi} \mathbf{E}^{i n c}- \boldsymbol{\chi} \mathbf{G}_D \boldsymbol{\Phi}_\theta(\gamma(\boldsymbol{\rho}))\right\|_2}{\left\|\boldsymbol{\chi} \mathbf{E}^{i n c}\right\|_2} \\ 
    &+\frac{\left\|\mathbf{d}^*-\left|\mathbf{E}_r^{i n c}+\mathbf{G}_S \boldsymbol{\Phi}_\theta(\gamma(\boldsymbol{\rho}))\right|\right\|_2}{\left\|\mathbf{d}^*\right\|_2}+\alpha \sum|\nabla \boldsymbol{\chi}|
\end{aligned}
\end{equation}
where $|\cdot|$ represents the magnitude operation. The measurement data $\mathbf{d}^* = |\mathbf{E}^{\mathrm{tot}}_{\mathrm{measured}}|$ denotes the total field magnitude recorded at the receivers, which is directly measurable in practice. The term $\mathbf{E}^{\mathrm{inc}}_r$ represents the incident field at the receiver locations, analytically computable from the known transmitter positions and background medium parameters. The magnitude operation is applied to the predicted total field $|\mathbf{E}^{\mathrm{inc}}_r + \mathbf{G}_S \boldsymbol{\omega}|$, so that no separation of the scattered field from the total field is needed. This formulation makes the phaseless inversion directly applicable to practical scenarios where only total field amplitudes are available \cite{chen2018computational}. Thanks to the differentiable formulation of DeepCSI, the full-data and phaseless-data cases are handled within the same framework by simply redefining the data-consistency term. Finally, our DeepCSI method is summarized in Algorithm~\ref{alg:deepcsi}.

\section{Numerical Results}
In this section, the performance of DeepCSI is evaluated through the inversion of both synthetic and experiment data, and compared with CSI \cite{van1997contrast}, MRCSI \cite{van2001contrast} methods, and additional unsupervised learning baselines. The experiments aim to verify the accuracy and robustness of the proposed method under various measurement conditions, including full-data, phaseless-data, and multi-frequency scenarios. Unless otherwise specified, the network and optimizer settings follow those described in Section III.

\begin{table}[t]

\centering
\caption{Hyperparameter settings used in all experiments.}
\label{hyper_setting}
\begin{tabular}{l>{\centering\arraybackslash}p{1.2cm}>{\centering\arraybackslash}p{1.5cm}}
\hline
\textbf{Parameter} & \textbf{Symbol} & \textbf{Value} \\ \hline
\multicolumn{3}{l}{\textit{Network architecture}} \\
\quad Positional encoding bandwidth & $M$ & 10 \\
\quad Residual blocks & $N_b$ & 3 \\
\quad Hidden width & -- & 256 \\
\quad Activation & -- & GELU \\
\quad Output channels & -- & 2 (Re, Im) \\ \hline
\multicolumn{3}{l}{\textit{Optimization}} \\
\quad ResMLP learning rate & $\eta_\theta$ & 0.05 \\
\quad Contrast learning rate & $\eta_\chi$ & 0.1 \\
\quad LR decay schedule & -- & $\times 0.1^{n/2000}$ \\
\quad Optimizer & -- & Adam \\
\quad Iterations & $N_{\mathrm{iter}}$ & 1000 \\ \hline
\multicolumn{3}{l}{\textit{Regularization}} \\
\quad TV norm & -- & L1 \\
\quad TV weight & $\alpha$ & 0.1 \\
\quad TV weight decay & -- & linear, 0.05 \\ \hline
\multicolumn{3}{l}{\textit{Problem setup}} \\
\quad Inversion grid & $N \times N$ & $64 \times 64$ \\
\quad Forward grid & -- & $128 \times 128$ \\ \hline
\end{tabular}
\end{table}

To quantitatively assess the accuracy of inversion, relative root mean square error (RMSE) and structural similarity index (SSIM) \cite{wang2004image} are used. Their formulas are as follows:
\begin{equation}
\operatorname{RMSE}=\frac{\left\|p_{\text {inv }}-p_{\text {truth }}\right\|_2}{\left\|p_{\text {truth }}\right\|_2}=\sqrt{\frac{\sum\left(p_{\text {inv }}-p_{\text {truth }}\right)^2}{\sum\left(p_{\text {truth }}\right)^2}}
\end{equation}
\begin{equation}
\operatorname{SSIM}=\frac{\left(2 \mu_{i n v} \mu_{t r u t h}+C_1\right)\left(2 \sigma_{i n v, t r u t h}+C_2\right)}{\left(\mu_{i n v}^2+\mu_{t r u t h}^2+C_1\right)\left(\sigma_{i n v}^2+\sigma_{t r u t h}^2+C_2\right)}
\end{equation}
where $p_{inv}$ and $p_{truth}$ represent the inversion results and the ground truth, respectively. $\mu_{inv}$ and $\mu_{truth}$ denote the mean values of $p_{inv}$ and $p_{truth}$, while $\sigma_{inv}$ and $\sigma_{truth}$ represent the corresponding variances. $\sigma_{inv,truth}$ indicates the covariance between the inversion result and the ground truth. $C_1=(K_1L)^2$ and $C_2=(K_2 L)^2$ are two small constants introduced to avoid division by zero. Among them, $K_1$ and $K_2$ are two hyperparameters, set to 0.01 and 0.03, respectively. $L$ represents the dynamic range of pixel values in the results, which is set to 1 in this paper.

The proposed DeepCSI method, as well as the baseline methods such as CSI and MRCSI \cite{van2001contrast}, are implemented using the PyTorch framework \cite{paszke2019pytorch} and executed on a single NVIDIA RTX 4090 GPU. To facilitate reproducibility, the key hyperparameters used throughout all experiments are consolidated in Table \ref{hyper_setting}. The asymmetric learning rates $\eta_\theta$ and $\eta_\chi$ reflect the different roles of the two variable groups. The contrast tensor is a direct pixel-wise representation with $N^2$ degrees of freedom and benefits from faster updates to quickly establish a coarse permittivity estimate, whereas the ResMLP defines a continuous mapping with an implicit smoothness bias that requires a smaller learning rate to preserve this regularization effect. These values are determined empirically. The exponential learning rate decay schedule further reduces sensitivity to the initial rates by annealing both toward zero during optimization.

\subsection{Synthetic Data Inversion}
\begin{figure}[t]
	\centering
	\includegraphics[width=0.4\textwidth]{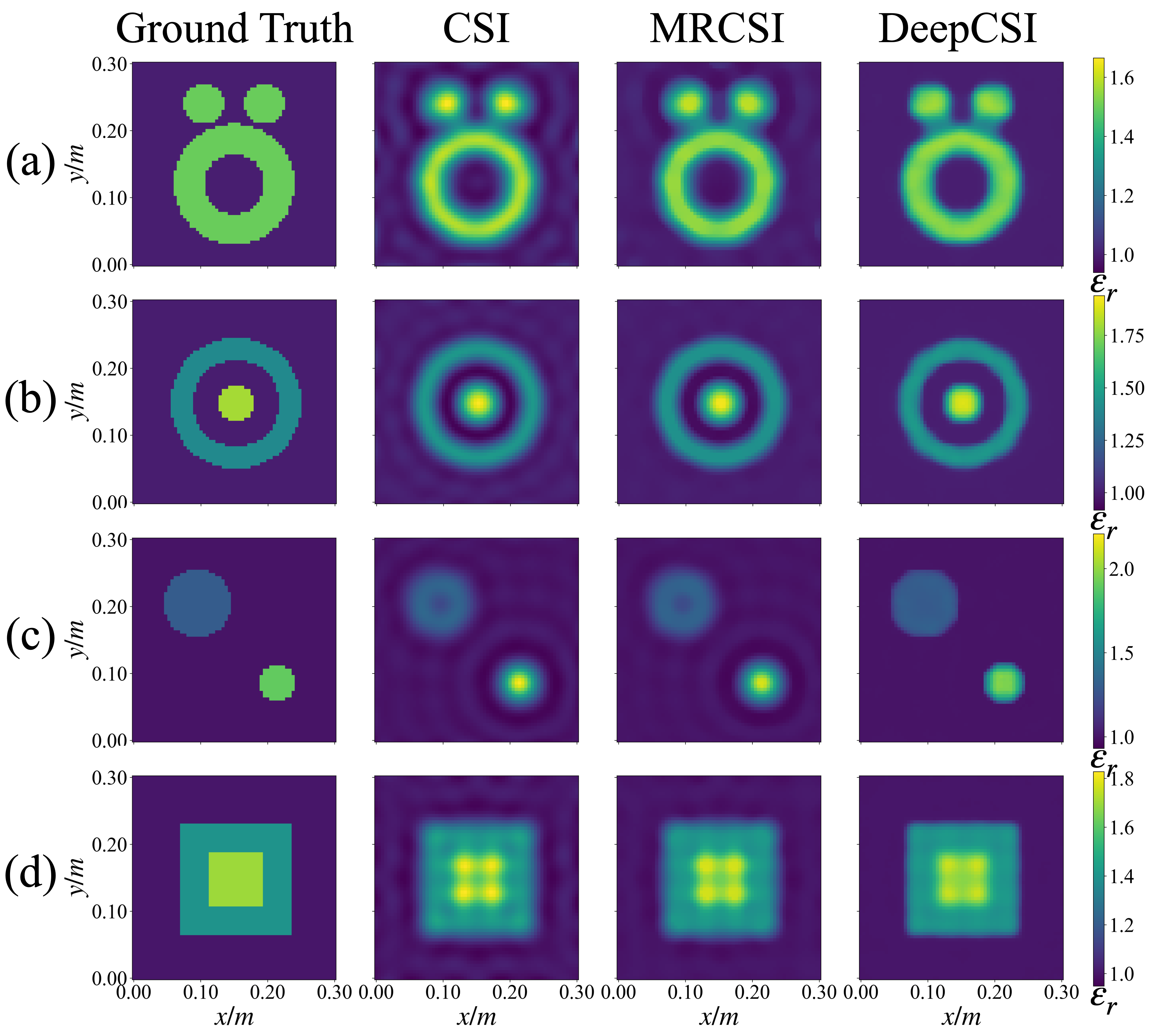}
	\caption{FD-SF inversion results for CSI, MRCSI, and DeepCSI at 3 GHz. The first column represents the ground truth, while columns 2 to 4 represent the inversion results of CSI, MRCSI, and DeepCSI, respectively. (a)$\sim$(d) display four representative test cases.}
	\label{figs_fdcomp}
\end{figure}

\begin{figure*}[t]
	\centering
	\includegraphics[width=0.9\textwidth]{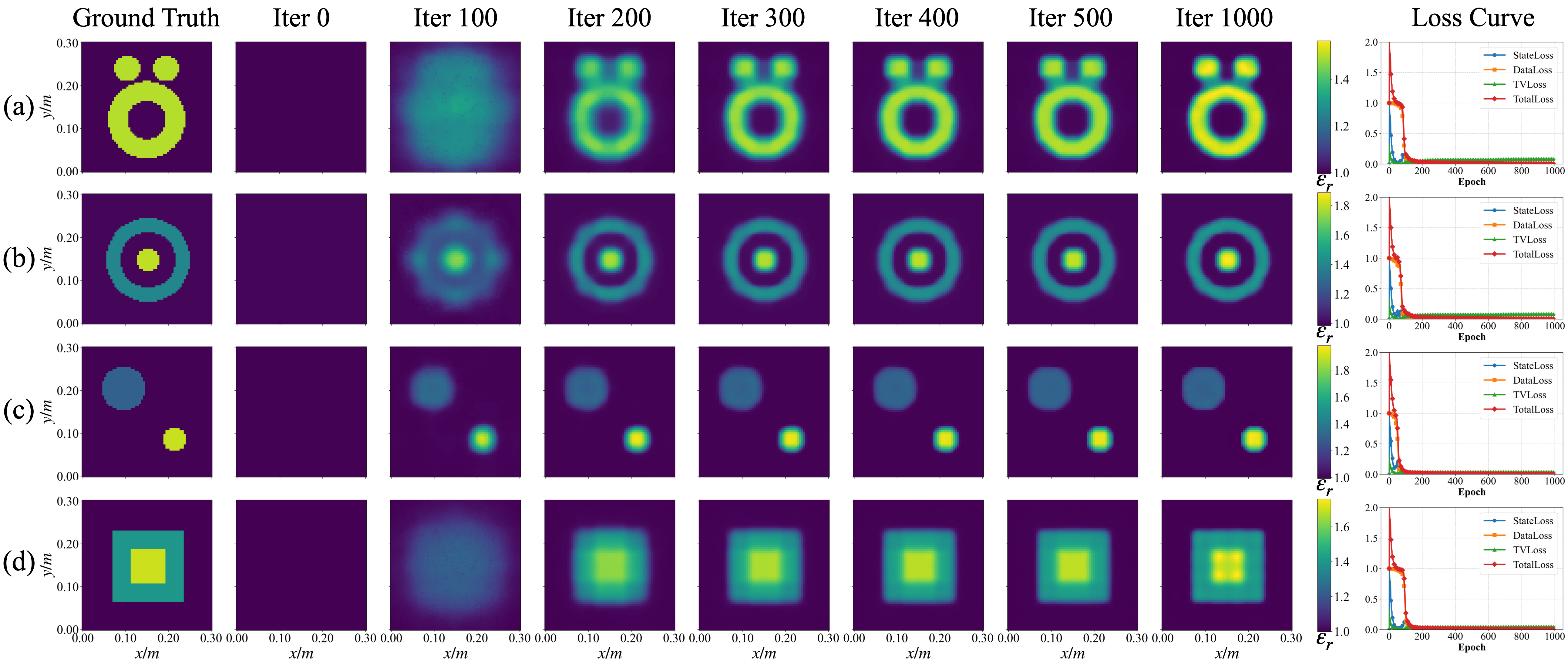}
	\caption{Iterative reconstruction process and loss curve of DeepCSI for FD-SF inversion at 3 GHz. The first column represents the ground truth, while columns 2 to 8, labeled with "Iter $n$", present the inversion results after the $n$-th iteration. The final column displays the loss curve of DeepCSI for the corresponding test case. (a)$\sim$(d) present four representative test cases.}
	\label{figs_fdsteps}
\end{figure*}

To evaluate the reconstruction capability of DeepCSI, synthetic data are generated and inverted under different measurement conditions. The DOI is set to 0.3$\times$0.3 $\text{m}^2$, with 16 transmitters and 32 receivers uniformly distributed on circles with radii of 2 m and 2.2 m, respectively. For the single-frequency (SF) inversion, the operating frequency is 3 GHz, while for the multi-frequency (MF) inversion, three frequencies of 3, 4, and 5 GHz are used. The field data used for inversion are obtained through Method of Moments (MoM) \cite{harrington1993field}. To avoid inverse crime, the DOI is discretized into 128$\times$128 grids in forward modeling and 64$\times$64 grids in inversion. Furthermore, Gaussian white noise with a specified level $\sigma_n$ is added to the scattered field data according to \cite{carpio2020bayesian}:
\begin{equation}
\label{eq_noise}
\hat{\mathbf{E}}_{s c a}=\mathbf{E}_{s c a}+\frac{\sigma_n * \mu_{sca}}{\sqrt{2}}(\boldsymbol{\mu}_1 + j \boldsymbol{\mu}_2)
\end{equation}
where $\boldsymbol{\mu}_1$ and $\boldsymbol{\mu}_2$ are independent Gaussian random vectors following $\mathcal{N}(0,1)$, $\mu_{sca}$ represents the average magnitude of the received scattered field. Unless otherwise specified, the noise level $\sigma_n$ is set to 5\%. The relationship between $\sigma_n$ and the signal-to-noise ratio (SNR) depends on the amplitude distribution of the scattered field. Under the noise model in Eq.~(20), the SNR can be approximately expressed as:
\begin{equation}
\mathrm{SNR_{\mathrm{dB}}} \approx 10\log_{10}\!\left(\frac{c}{\sigma_n^2}\right)
\end{equation}
where $c = \mathrm{mean}(|E^{\mathrm{sca}}|^2)\,/\,[\mathrm{mean}(|E^{\mathrm{sca}}|)]^2$ is determined by the amplitude distribution of the scattered field. For the test cases in this paper, $c \approx 2.2$--$3.0$, corresponding to approximately 30~dB at $\sigma_n = 5\%$ and 4~dB at $\sigma_n = 100\%$. Although such a severe noise level is uncommon in well-controlled laboratory measurements, it is included here to probe the robustness limit of the proposed method under extremely adverse sensing conditions.

\subsubsection{Full Data Inversion}
The full-data single-frequency (FD-SF) inversion results at 3 GHz are presented in Fig. \ref{figs_fdcomp} and Table \ref{tab_fd_comp}. For the four representative test cases, the inversion results obtained using the proposed DeepCSI method not only yield a contrast and shape closer to the true values, but also exhibit fewer artifacts in the background region. The quantitative comparisons in Table \ref{tab_fd_comp} further demonstrate that, in all four cases, the proposed method achieves lower RMSE and higher SSIM. These results indicate that the proposed method improves inversion accuracy compared to traditional CSI and MRCSI methods.

The iterative reconstruction process of DeepCSI is shown in Fig. \ref{figs_fdsteps}. It can be seen that starting from a uniform initial background, DeepCSI requires only 200 Adam iterations to obtain a coarse estimation, and after 500 iterations, the result closely matches the ground truth. Considering that each update requires approximately 0.05 seconds on the GPU platform, the proposed algorithm achieves satisfactory inversion results in 10 to 25 seconds, depending on the convergence rate of each case. Under the same hardware platform, CSI and MRCSI methods require about 150 iterations to converge and take approximately 30 seconds in total. These comparisons indicate that DeepCSI achieves higher inversion accuracy while maintaining comparable computational efficiency, largely because the neural implicit parameterization enables more faithful modeling of the contrast source.

\begin{figure}[t]
	\centering
	\includegraphics[width=0.4\textwidth]{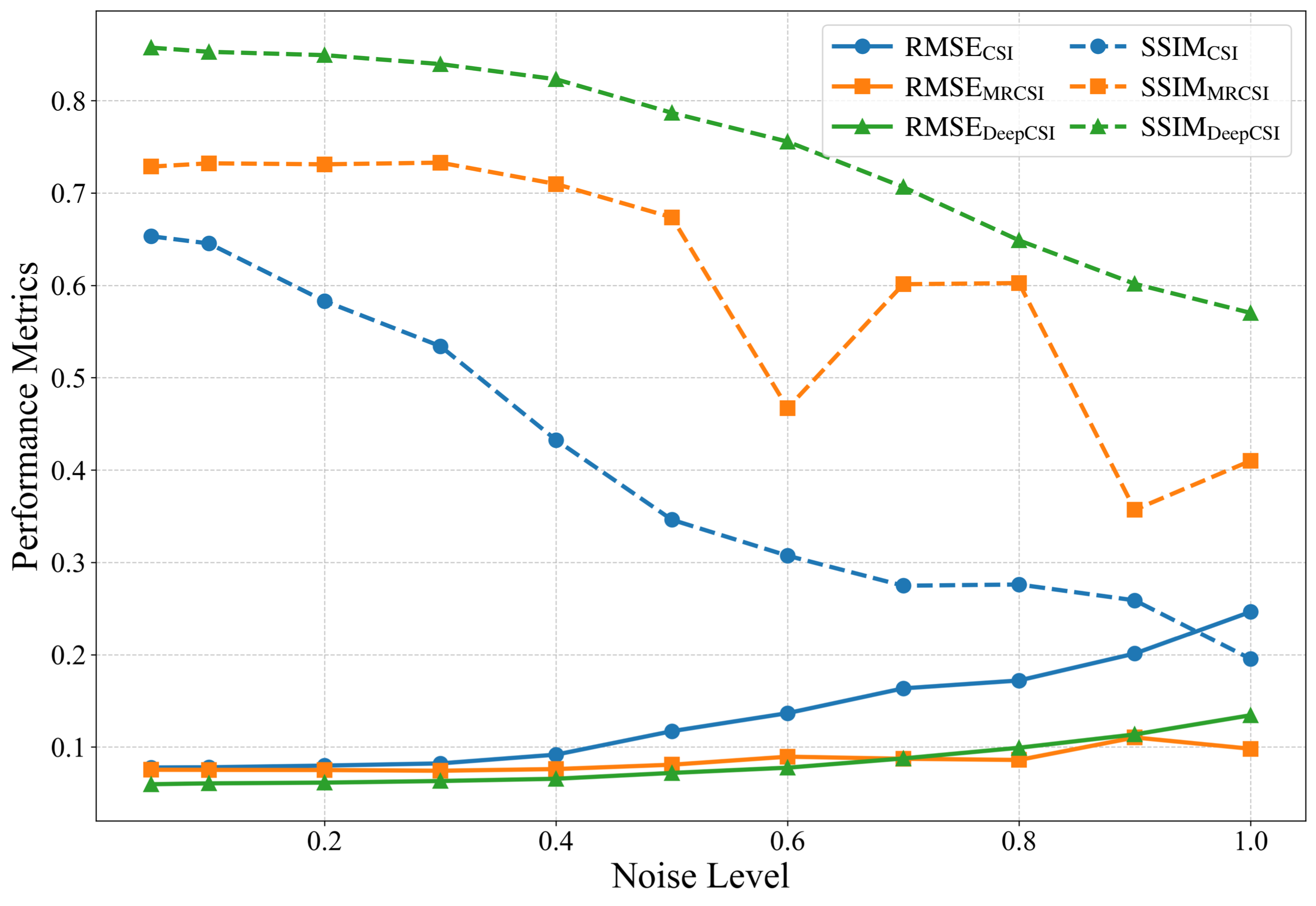}
	\caption{RMSE and SSIM of FD-SF inversion results obtained by CSI, MRCSI, and DeepCSI under various noise levels.}
	\label{figs_fdsnrcurve}
\end{figure}

In practical measurements, noise interference in the data is inevitable, and therefore, inversion algorithms are usually required to have good noise resistance. To evaluate the performance of DeepCSI, Gaussian white noise with levels ranging from 5 \% to 100 \% is added to the scattered-field data of the Austria model. The RMSE and SSIM of the inversion results obtained by CSI, MRCSI and DeepCSI under various noise levels are illustrated in Fig. \ref{figs_fdsnrcurve}. As the noise level increases, the inversion accuracy of all methods gradually decreases. However, DeepCSI consistently outperforms CSI and MRCSI, particularly in SSIM.
\begin{table}[t]
	\centering
	\caption{FD-SF inversion performance of CSI, MRCSI and DeepCSI on four representative cases at 3~GHz}
	\label{tab_fd_comp}
	\begin{tabular}{c|>{\centering\arraybackslash}p{1.5cm}>{\centering\arraybackslash}p{1.2cm}>{\centering\arraybackslash}p{1.2cm}}
		\hline
		Case No.           & Method  & RMSE$\downarrow$   & SSIM$\uparrow$   \\ \hline
		\multirow{3}{*}{(a)} & CSI     & 0.0778 & 0.6531 \\
		& MRCSI   & 0.0754 & 0.7285 \\
		& DeepCSI & \textbf{0.0597} & \textbf{0.8573} \\ \hline
		\multirow{3}{*}{(b)} & CSI     & 0.0665 & 0.7993 \\
		& MRCSI   & 0.0637 & 0.8598 \\
		& DeepCSI & \textbf{0.0532} & \textbf{0.9112} \\ \hline
		\multirow{3}{*}{(c)} & CSI     & 0.0644 & 0.7770 \\
		& MRCSI   & 0.0631 & 0.8239 \\
		& DeepCSI & \textbf{0.0410} & \textbf{0.9610} \\ \hline
		\multirow{3}{*}{(d)} & CSI     & 0.0493 & 0.7406 \\
		& MRCSI   & 0.0467 & 0.8102 \\
		& DeepCSI & \textbf{0.0331} & \textbf{0.9098} \\ \hline
	\end{tabular}
	\vspace{-0.5em}
\end{table}

\begin{figure}[t]
	\centering
	\includegraphics[width=0.47\textwidth]{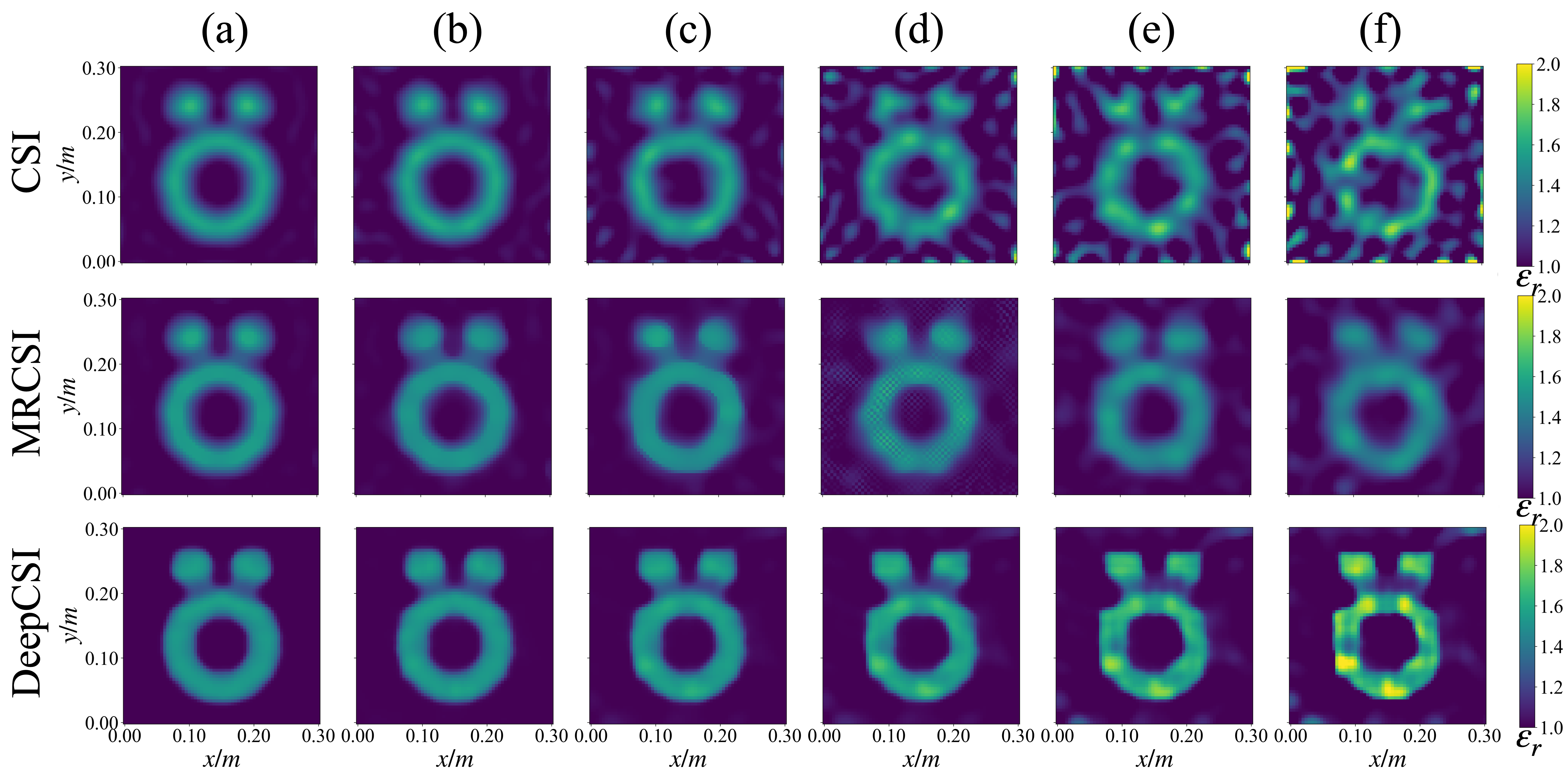}
	\caption{FD-SF inversion results for the Austria model using CSI, MRCSI, and DeepCSI under different noise levels. (a)$\sim$(f) represent the inversion results at noise levels of 5\%, 20\%, 40\%, 60\%, 80\%, and 100\%, respectively.}
	\label{figs_fdsnrcases}
\end{figure}

\begin{figure}[t]
	\centering
	\includegraphics[width=0.4\textwidth]{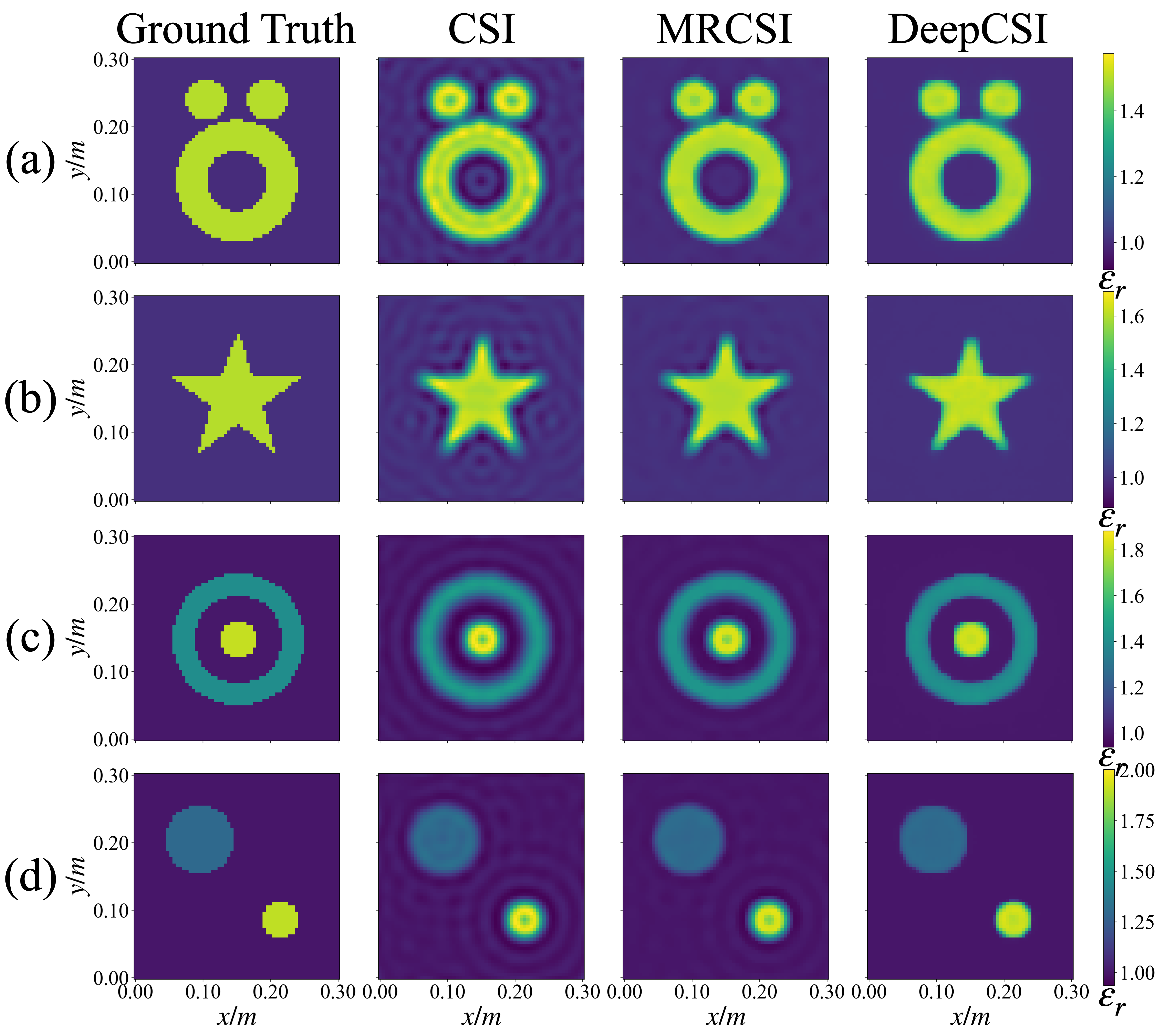}
	\caption{FD-MF inversion results for CSI, MRCSI, and DeepCSI at 3, 4, and 5 GHz. The first column represents the ground truth, while columns 2 to 4 represent the inversion results of CSI, MRCSI, and DeepCSI, respectively. (a)$\sim$(d) display four representative test cases.}
	\label{figs_multiinv}
\end{figure}

Fig. \ref{figs_fdsnrcases} compares the inversion results of CSI, MRCSI and DeepCSI under different noise levels. When the noise level exceeds 40\%, the results from CSI and MRCSI begin to exhibit pronounced artifacts, and the target boundaries gradually become blurred as the noise level increases. In contrast, the inversion results of DeepCSI maintain a clean background, with the target boundary remaining clear. Although there are some deviations in the inverted medium parameters, the overall inversion accuracy still outperforms CSI and MRCSI, demonstrating strong noise resistance. This improvement is attributed to the neural contrast-source parameterization, which introduces implicit regularization and thereby suppresses noise-induced nonphysical artifacts. It is also worth noting that although the noise level is 100\%, the scattering field data are not completely overwhelmed by the noise, as indicated in Eq. \eqref{eq_noise}. Instead, some information with amplitudes higher than the average is retained, enabling the target reconstruction. 

The effectiveness of full-data multi-frequency (FD-MF) inversion is further validated at 3, 4, and 5 GHz, as shown in Fig. \ref{figs_multiinv}, and the corresponding performance metrics are presented in Table \ref{tab_mf_comp}.

As shown in Table \ref{tab_mf_comp}, compared with the single-frequency (SF) inversion results in Table \ref{tab_fd_comp}, the RMSE from all three methods decreases and the SSIM improves under MF measurements. DeepCSI still achieves higher accuracy in MF inversion compared to CSI and MRCSI. Furthermore, as seen in Fig. \ref{figs_multiinv}, DeepCSI provides the most accurate and physically consistent reconstructions among all methods. In particular, the recovered ring-type structures in cases (a) and (c) exhibit thicknesses that closely match the ground truth, verifying that DeepCSI can effectively exploit spectral diversity to enhance inversion accuracy and stability.
\begin{table}[t]
	\centering
	\caption{FD-MF inversion performance of CSI, MRCSI, and DeepCSI on four representative cases}
	\label{tab_mf_comp}
	\begin{tabular}{c|>{\centering\arraybackslash}p{1.5cm}>{\centering\arraybackslash}p{1.2cm}>{\centering\arraybackslash}p{1.2cm}}
		\hline
		Case No.           & Method  & RMSE$\downarrow$   & SSIM$\uparrow$   \\ \hline
		\multirow{3}{*}{(a)} & CSI     & 0.0586 & 0.7885 \\
		& MRCSI   & 0.0550 & 0.8734 \\
		& DeepCSI & \textbf{0.0452} & \textbf{0.9235} \\ \hline
		\multirow{3}{*}{(b)} & CSI     & 0.0570 & 0.8046 \\
		& MRCSI   & 0.0543 & 0.8979 \\
		& DeepCSI & \textbf{0.0429} & \textbf{0.9471} \\ \hline
		\multirow{3}{*}{(c)} & CSI     & 0.0599 & 0.7965 \\
		& MRCSI   & 0.0559 & 0.8759 \\
		& DeepCSI & \textbf{0.0357} & \textbf{0.9627} \\ \hline
		\multirow{3}{*}{(d)} & CSI     & 0.0477 & 0.8373 \\
		& MRCSI   & 0.0453 & 0.9168 \\
		& DeepCSI & \textbf{0.0285} & \textbf{0.9829} \\ \hline
	\end{tabular}
\end{table}

\begin{figure}[t]
	\centering
	\includegraphics[width=0.35\textwidth]{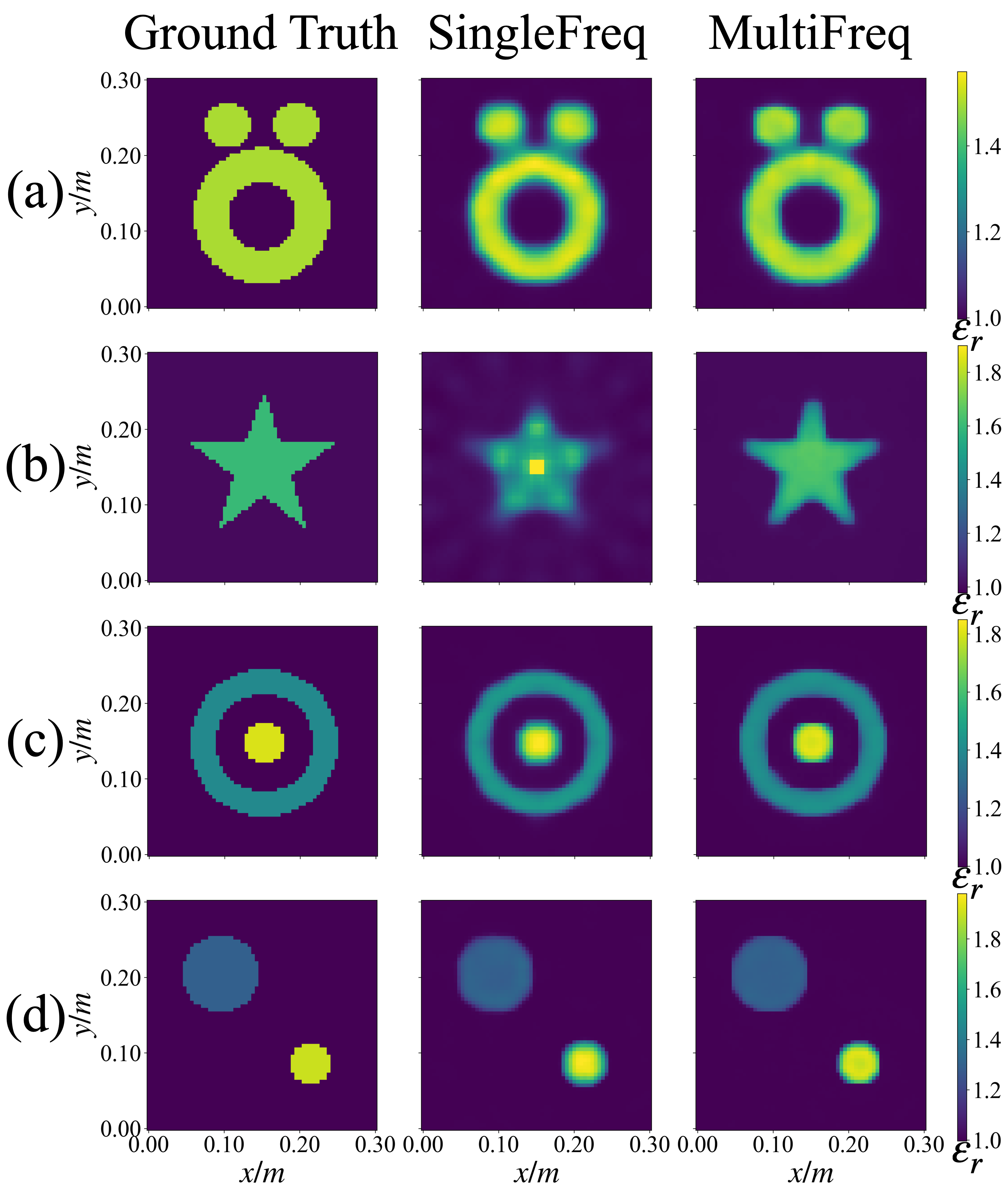}
	\caption{PD inversion results of DeepCSI under SF and MF measurements. The first column represents the ground truth, while columns 2 to 3 represent the inversion results of DeepCSI under SF and MF measurements, respectively. (a)$\sim$(d) display four representative test cases.}
	\label{figs_pdcomp}
\end{figure}

\begin{table}[t]
	\centering
	\caption{PD inversion performance of DeepCSI under SF and MF measurements on four representative cases}
	\label{tab_pd_comp}
	\begin{tabular}{c|>{\centering\arraybackslash}p{1.8cm}>{\centering\arraybackslash}p{1.2cm}>{\centering\arraybackslash}p{1.2cm}}
		\hline
		Case No.           & Method  & RMSE$\downarrow$   & SSIM$\uparrow$   \\ \hline
		\multirow{2}{*}{(a)} & SF-DeepCSI     & 0.0613 & 0.8497 \\
		& MF-DeepCSI & \textbf{0.0490} & \textbf{0.9090} \\ \hline
		\multirow{2}{*}{(b)} & SF-DeepCSI     & 0.3153 & 0.7847 \\
		& MF-DeepCSI & \textbf{0.0443} & \textbf{0.9425} \\ \hline
		\multirow{2}{*}{(c)} & SF-DeepCSI     & 0.0557 & 0.9058 \\
		& MF-DeepCSI & \textbf{0.0373} & \textbf{0.9571} \\ \hline
		\multirow{2}{*}{(d)} & SF-DeepCSI     & 0.0417 & 0.9599 \\
		& MF-DeepCSI & \textbf{0.0301} & \textbf{0.9802} \\ \hline
	\end{tabular}
	\vspace{-0.5em}
\end{table}

\subsubsection{Phaseless Data Inversion}
DeepCSI can be easily extended to invert phaseless data. Given that MF data are often used in phaseless measurements, this section applies DeepCSI to invert both SF scattered field data at 3 GHz and MF scattered field data at 3, 4, and 5 GHz. The results are shown in Fig. \ref{figs_pdcomp}, and the reconstruction accuracy is quantified and presented in Table \ref{tab_pd_comp}.

\begin{figure}[t]
	\centering
	\includegraphics[width=0.4\textwidth]{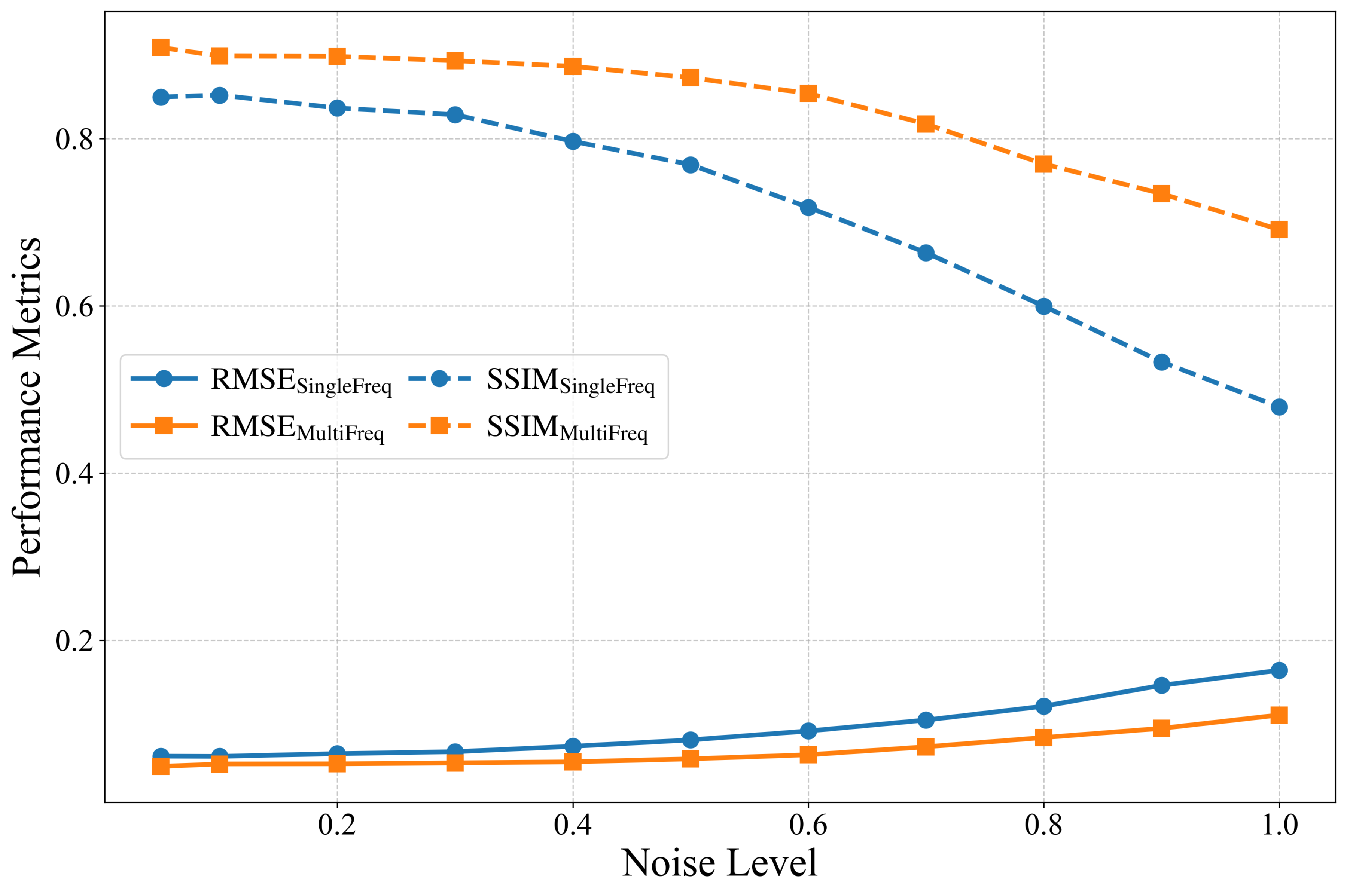}
	\caption{RMSE and SSIM of PD inversion results obtained by DeepCSI under SF and MF measurements at different noise levels.}
	\label{figs_pdsnrcurve}
\end{figure}
\begin{figure}[!t]
	\centering
	\includegraphics[width=0.5\textwidth]{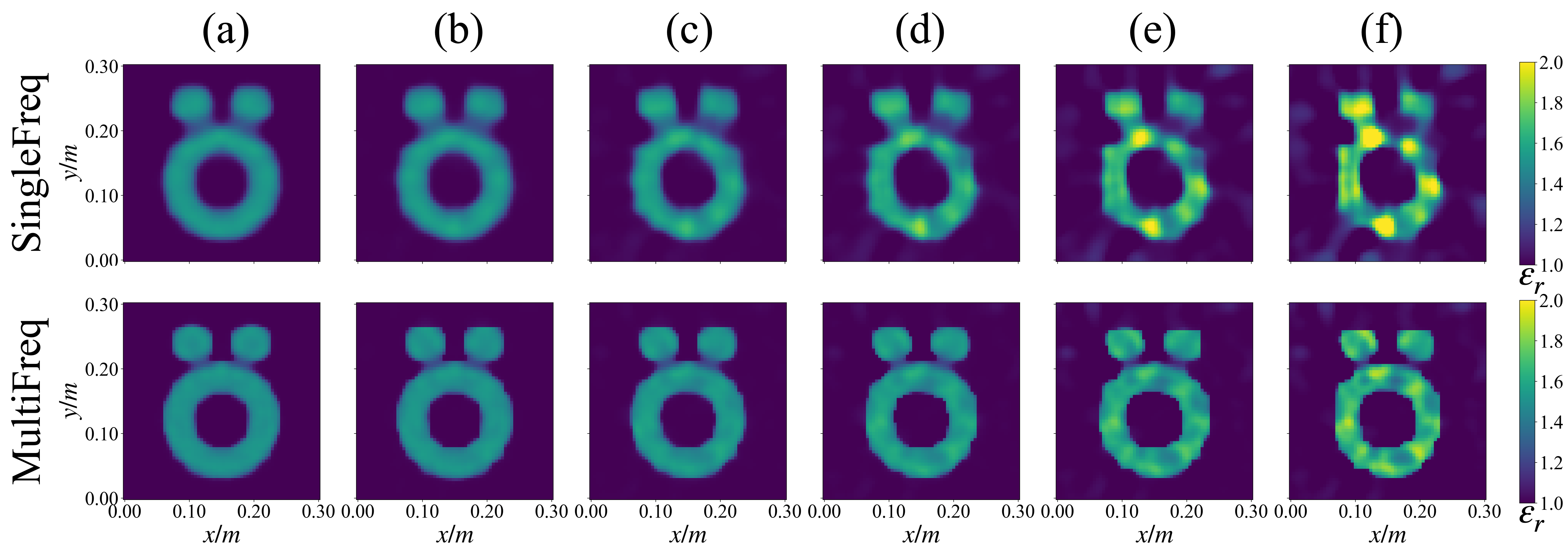}
	\caption{PD inversion results of DeepCSI for Austria model under SF and MF measurements at different noise levels. (a)$\sim$(f) represent the inversion results at noise levels of 5\%, 20\%, 40\%, 60\%, 80\%, and 100\%, respectively.}
	\label{figs_pdsnrcases}
\end{figure}

As shown in Fig. \ref{figs_pdcomp} and Table \ref{tab_pd_comp}, DeepCSI can reconstruct targets using SF phaseless measurement. In most cases, the reconstructed target shape, size, and permittivity are accurate. Only in case (b), the inversion result of DeepCSI deviates from the ground truth, though they still reflect the shape and permittivity of the target. Furthermore, when using MF phaseless data, the inversion results of DeepCSI show more accurate shapes, medium parameters and clearer edges. This indicates that DeepCSI effectively exploits phaseless measurements and yields accurate reconstructions under both single- and multi-frequency settings.

Furthermore, to evaluate the noise robustness of DeepCSI under phaseless conditions, Gaussian white noise at 11 different levels, ranging from 5\% to 100\%, is added to the total field data. The inversion of noisy phaseless total field data using the proposed method under both SF and MF conditions is performed, and the resulting RMSE and SSIM as functions of noise levels are presented in Fig. \ref{figs_pdsnrcurve}. As shown in the figure, with the increase of noise levels, the accuracy of inversion decreases in both cases. However, even with 100\% Gaussian white noise added, the RMSE of the SF inversion results from the proposed method remains below 0.2, with an SSIM close to 0.5, indicating that the inversion results still basically reflect the target information. Additionally, it can be observed that the RMSE and SSIM of the MF measurement inversion results are consistently better than those of the SF measurements, highlighting the advantages and necessity of MF data in phaseless measurements. 

To further illustrate the impact of noise on DeepCSI phaseless inversion, the inversion results for the Austria target under SF and MF measurements at different noise levels are visualized in Fig. \ref{figs_pdsnrcases}. As the noise level increases, the target shape of SF inversion gradually deviates from the ground truth, and the accuracy of the permittivity inversion decreases. However, even with 100\% noise added, the SF reconstruction results can still approximate the shape of the target. Furthermore, it can be observed that the target shape, size, and permittivity in MF inversion results are closer to the ground truth. These results show that DeepCSI can effectively utilize phaseless measurement data for target inversion while remaining robust to noise.

\subsubsection{Effect of Neural Representation and Regularization}
 
To clarify the contributions of the neural contrast source representation and the explicit regularization scheme, a factorial ablation study is conducted with six configurations. Two representation types, namely pixel-based with the CG optimizer and neural with Adam, are crossed with three regularization settings, namely no regularization, add. TV-L1 with $\alpha = 0.1$, and mult. TV. All experiments use the full-data single-frequency setting at 3~GHz with 5\% noise.

\begin{table}[t]
    \centering
    
    \caption{Ablation results of two representation types and three regularization schemes under FD-SF inversion at 3~GHz}
    \label{tab_reg}
    \begin{tabular}{c|>{\centering\arraybackslash}p{1.8cm}>{\centering\arraybackslash}p{1.1cm}>{\centering\arraybackslash}p{0.8cm}>{\centering\arraybackslash}p{0.8cm}}
    \hline
    Method & Repr. & Reg. & RMSE$\downarrow$ & SSIM$\uparrow$ \\ \hline
    CSI        & Pixel (CG)    & None              & 0.0668 & 0.812 \\
    CSI+TV  & Pixel (CG)    & Add. TV    & 0.0666 & 0.881 \\
    MRCSI      & Pixel (CG)    & Mult. TV & 0.0641 & 0.867 \\ \hline
    DeepCSI w/o Reg   & Neural (Adam) & None              & 0.0529 & 0.884 \\
    DeepCSI & Neural (Adam) & Add. TV    & \textbf{0.0481} & \textbf{0.910} \\
    DeepCSI w/ MR & Neural (Adam) & Mult. TV & 0.1523 & 0.621 \\ \hline
    \end{tabular}
\end{table}

\begin{figure}[t]
    \centering
    \includegraphics[width=\linewidth]{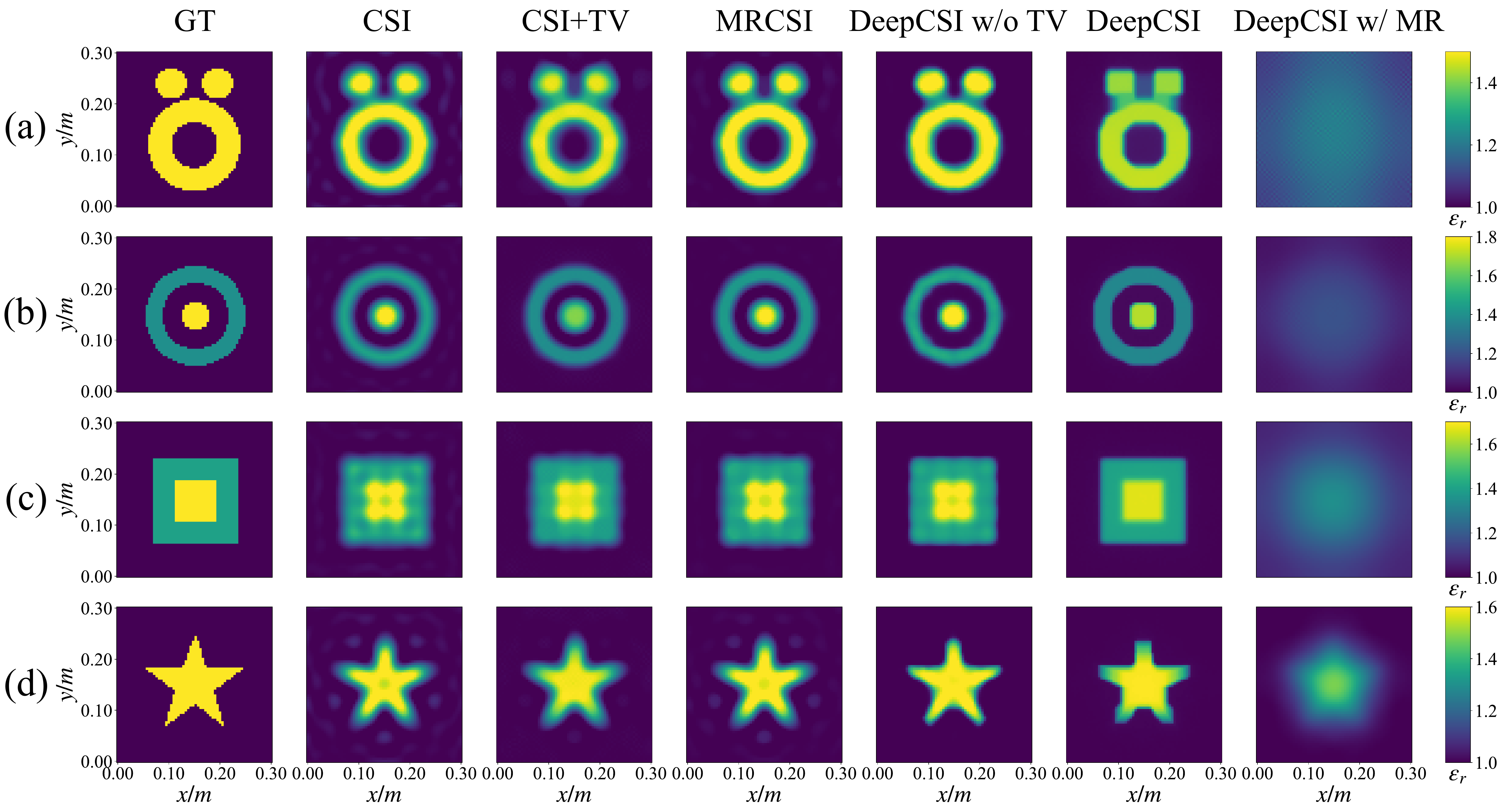}
    \caption{Reconstructed permittivity using six representation-regularization configurations across four representative cases under FD-SF inversion at 3~GHz. The columns from left to right represent ground truth, CSI, CSI+TV, MRCSI, DeepCSI without TV, DeepCSI, and DeepCSI with multiplicative regularization, respectively.}
    \label{fig_reg}
\end{figure}
    
Table~\ref{tab_reg} reports the average RMSE and SSIM over the four presentative cases, and Fig.~\ref{fig_reg} shows the corresponding reconstructions. The neural representation is the dominant factor in artifact suppression. DeepCSI without any regularization achieves an RMSE of 0.0529 and SSIM of 0.884, outperforming all pixel-based variants including MRCSI at 0.0641 and 0.867. Adding TV-L1 regularization provides a complementary but secondary benefit, further improving the metrics to 0.0481 and 0.910. The MRCSI-style mult. TV regularization fails when combined with the Adam optimizer, degrading RMSE to 0.1523 and SSIM to 0.621, because its ratio-based weighting mechanism is designed for the large, discrete updates of the CG line search and becomes numerically unstable under Adam's incremental updates. This confirms that the artifact suppression observed in DeepCSI originates primarily from the neural parameterization rather than from the regularization scheme.

\subsubsection{Comparison with Unsupervised Learning Baselines}
To validate the design choices underlying DeepCSI, we compare it with two unsupervised learning baselines under the same full-data single-frequency inversion setting. Together, these baselines allow us to examine the respective roles of the VIE-based physical constraint and the direct contrast-source parameterization.

The first baseline, referred to as Neural-$E^{\mathrm{tot}}$, uses the same ResMLP architecture to parameterize the total field~$E^{\mathrm{tot}}$ instead of the contrast source~$\omega$. The contrast source is then obtained by Eq.~\eqref{eq:omega_chi} during the forward pass, while all other settings remain identical to DeepCSI.

The second baseline, referred to as PINN-Helmholtz, replaces the VIE state equation with the Helmholtz PDE residual. Specifically, the total loss is defined as
\begin{equation}
\mathcal{L}_{\mathrm{PINN}} = \lambda_1 \mathcal{L}_{\mathrm{PDE}} + \lambda_2 \mathcal{L}_{\mathrm{data}} + \alpha \mathcal{L}_{\mathrm{TV}}
\end{equation}
where the PDE residual term is
\begin{equation}
\mathcal{L}_{\mathrm{PDE}} = \frac{1}{N^2} \sum_{i} \left| \nabla^2 E^{\mathrm{tot}}(\mathbf{r}_i) + k_0^2 \varepsilon_r(\mathbf{r}_i) E^{\mathrm{tot}}(\mathbf{r}_i) \right|^2
\end{equation}
and $\mathcal{L}_{\mathrm{data}}$ and $\mathcal{L}_{\mathrm{TV}}$ retain the same definitions as in DeepCSI. The Laplacian $\nabla^2 E^{\mathrm{tot}}$ is computed via second-order automatic differentiation at all DOI grid points. The loss weights $\lambda_1 = 0.001$ and $\lambda_2 = 1$ are selected from a grid search. To provide a favorable starting point, the permittivity is initialized with a backpropagation result rather than the free-space initialization used by DeepCSI and Neural-$E^{\mathrm{tot}}$. All other settings remain identical to DeepCSI.

\begin{table}[t]
\centering

\caption{FD-SF inversion performances of DeepCSI and two unsupervised learning baselines on four representative cases.}
\label{tab_baselines}
\begin{tabular}{c|>{\centering\arraybackslash}p{2.5cm}>{\centering\arraybackslash}p{1.0cm}>{\centering\arraybackslash}p{1.0cm}}
\hline
Case No. & Method & RMSE$\downarrow$ & SSIM$\uparrow$ \\ \hline
\multirow{3}{*}{(a)} & PINN-Helmholtz & 0.159 & 0.578 \\
& Neural-$E^{\mathrm{tot}}$ & 0.150 & 0.823 \\
& DeepCSI & \textbf{0.070} & \textbf{0.848} \\ \hline
\multirow{3}{*}{(b)} & PINN-Helmholtz & 0.140 & 0.638 \\
& Neural-$E^{\mathrm{tot}}$ & 0.099 & 0.877 \\
& DeepCSI & \textbf{0.048} & \textbf{0.936} \\ \hline
\multirow{3}{*}{(c)} & PINN-Helmholtz & 0.147 & 0.784 \\
& Neural-$E^{\mathrm{tot}}$ & 0.109 & 0.952 \\
& DeepCSI & \textbf{0.046} & \textbf{0.956} \\ \hline
\multirow{3}{*}{(d)} & PINN-Helmholtz & 0.142 & 0.755 \\
& Neural-$E^{\mathrm{tot}}$ & 0.091 & 0.888 \\
& DeepCSI & \textbf{0.064} & \textbf{0.892} \\ \hline
\end{tabular}
\end{table}

\begin{figure}[t]
\centering
\includegraphics[width=0.85\linewidth]{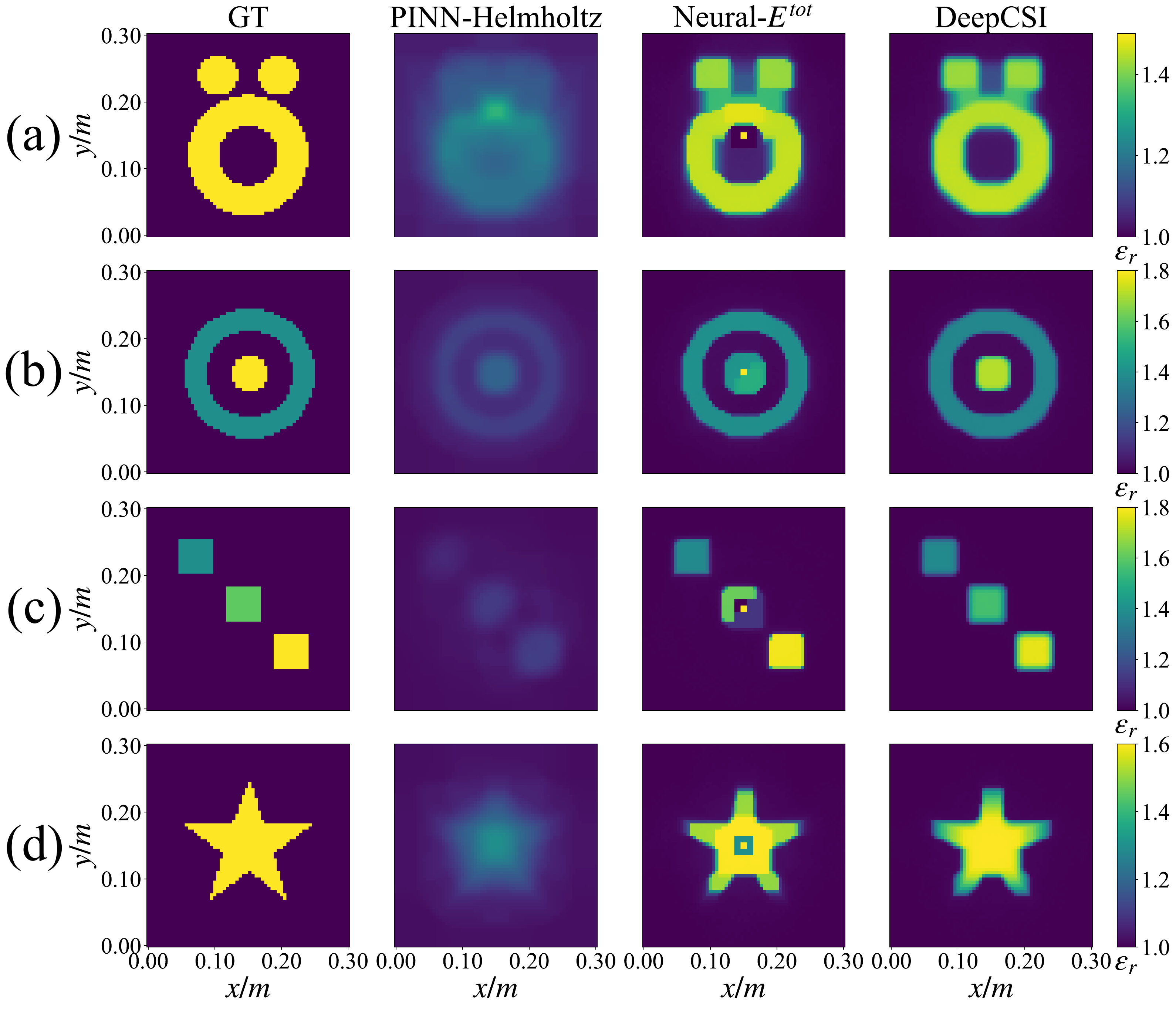}
\caption{FD inversion results of three unsupervised learning methods at 3 GHz on four representative cases. The columns from left to right correspond to ground truth, PINN-Helmholtz, Neural-$E^{\mathrm{tot}}$, and DeepCSI, respectively.}
\label{fig_baselines}
\end{figure}
 
Table~\ref{tab_baselines} and Fig.~\ref{fig_baselines} present the quantitative and visual results, respectively. The two baselines isolate the two core design choices of DeepCSI. PINN-Helmholtz achieves an average RMSE of only 0.147 and produces blurred, low-contrast reconstructions across all test cases. This is because the pointwise PDE residual used in PINN only imposes local constraints, whereas the VIE explicitly encodes global wave coupling through the Green's function convolution $\mathbf{G}_D \omega$. Neural-$E^{\mathrm{tot}}$ retains the VIE formulation and achieves an average RMSE of 0.112, recovering the correct spatial structure in most cases. However, this baseline exhibits unstable permittivity values due to the non-uniqueness of the factorization of Eq.~\eqref{eq:omega_chi}, where a multiplicative factor in~$\chi$ can be compensated by an inverse factor in~$E^{\mathrm{tot}}$ without affecting the data fit. DeepCSI avoids this non-uniqueness through direct $\omega$-parameterization and achieves the best average RMSE of 0.057 and SSIM of 0.907.

\subsubsection{Super-Resolution Verification}

A key advantage of the continuous neural representation is that it can be evaluated at spatial coordinates finer than the training grid. To validate this capability under a stringent setting, DeepCSI is trained on a coarse $32 \times 32$ grid at 5~GHz, rather than on the standard $64 \times 64$ inversion grid, with the positional encoding bandwidth set to $M = 9$ to match the coarser grid density. The trained ResMLP is then queried at $128 \times 128$ and $256 \times 256$ resolutions. The continuous output, referred to as DeepCSI-Query, is compared against bilinear interpolation of the coarse-grid result, referred to as DeepCSI-Interp., using the forward-solved contrast source at the target resolution as ground truth (GT).

\begin{table}[t]
\centering

\caption{Contrast source super-resolution results on the Austria model at 5~GHz}
\label{tab_cs_superres}
\begin{tabular}{c|>{\centering\arraybackslash}p{1.0cm}>{\centering\arraybackslash}p{1.0cm}|>{\centering\arraybackslash}p{1.0cm}>{\centering\arraybackslash}p{1.0cm}}
\hline
Target & \multicolumn{2}{c|}{DeepCSI-Interp.} & \multicolumn{2}{c}{DeepCSI-Query} \\
resolution & RMSE & SSIM & RMSE & SSIM \\ \hline
$128 \times 128$ & 0.3487 & 0.8012 & \textbf{0.2926} & \textbf{0.8312} \\
$256 \times 256$ & 0.3563 & 0.7736 & \textbf{0.2925} & \textbf{0.8015} \\ \hline
\end{tabular}
\end{table}

\begin{figure}[t]
\centering
\includegraphics[width=0.8\linewidth]{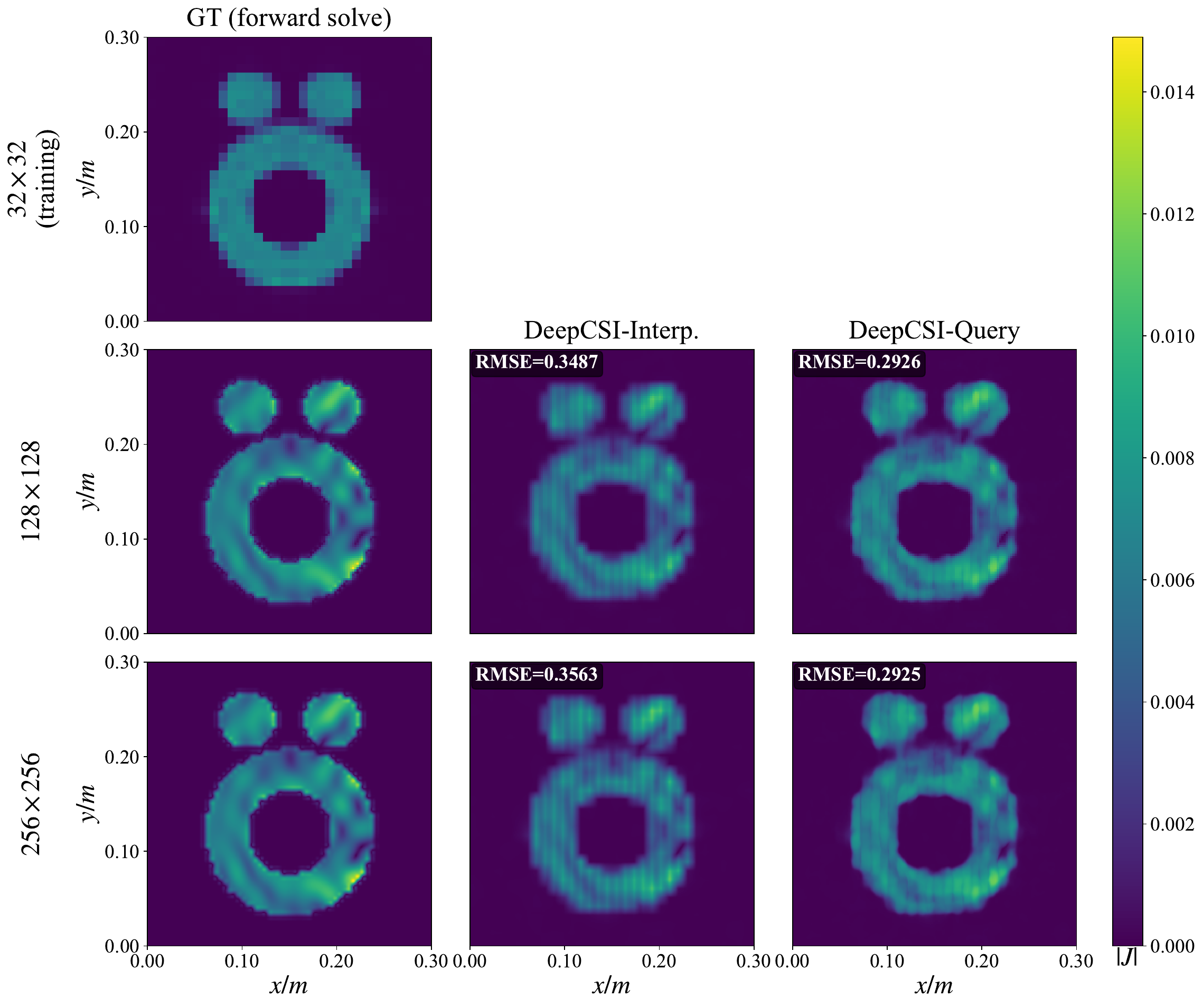}
\caption{Super-resolution results of the contrast source on the Austria model at 5~GHz. DeepCSI is trained on a $32 \times 32$ grid and evaluated at $128 \times 128$ and $256 \times 256$. Columns from left to right correspond to GT, DeepCSI-Interp., and DeepCSI-Query, respectively.}
\label{fig_cs_superres}
\end{figure}

As shown in Table~\ref{tab_cs_superres} and Fig.~\ref{fig_cs_superres}, DeepCSI-Query consistently outperforms DeepCSI-Interp. by 16--18\% in RMSE across all target resolutions, confirming that the network learns meaningful continuous structure beyond the discrete training points. The query error remains nearly constant as the target resolution increases, from 0.2926 at $128 \times 128$ to 0.2925 at $256 \times 256$, while the interpolation error grows from 0.3487 to 0.3563 over the same range, demonstrating resolution-independent generalization.

\begin{table}[t]
    \centering
    
    \caption{Reconstruction performance of four methods at $128 \times 128$ resolution on the Austria model at 5~GHz}
    \label{tab_contrast_superres}
    \begin{tabular}{c|>{\centering\arraybackslash}p{1.3cm}>{\centering\arraybackslash}p{1.0cm}>{\centering\arraybackslash}p{1.0cm}>{\centering\arraybackslash}p{1.0cm}}
    \hline
    Method & Inv. Grid & RMSE$\downarrow$ & SSIM$\uparrow$ & Time (s) \\ \hline
    CSI-TV & $128\times128$ & 0.0575 & 0.8051 & 3234 \\
    MRCSI & $128\times128$ & 0.0648 & 0.6893 & 1547 \\
    DeepCSI-Interp. & $32\times32$ & 0.0589 & 0.8284 & ${\sim}22$ \\ 
    DeepCSI-Query & $32\times32$ & \textbf{0.0469} & \textbf{0.8849} & ${\sim}\textbf{23}$ \\ \hline
    \end{tabular}
\end{table}

\begin{figure}[t]
    \centering
    \includegraphics[width=\linewidth]{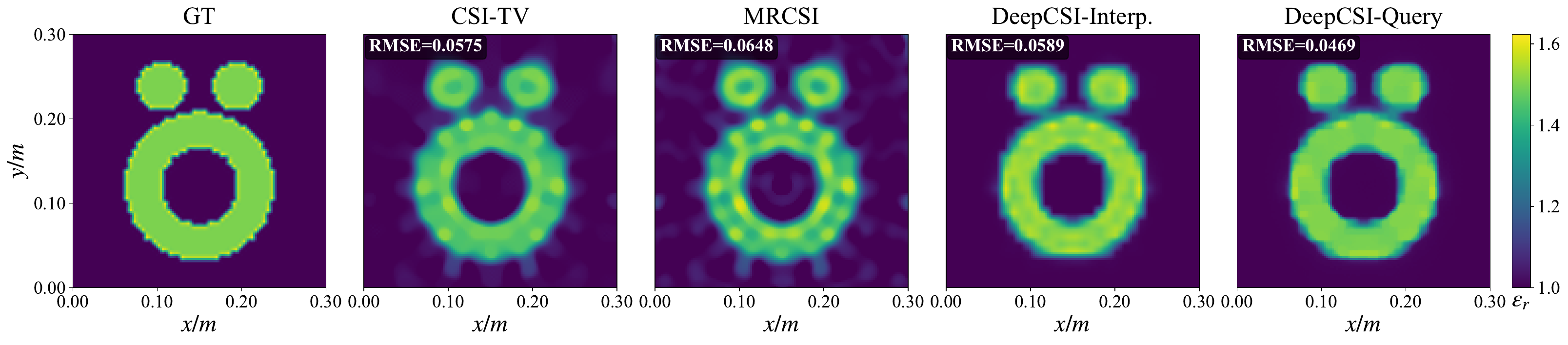}
    \caption{Reconstruction results at $128 \times 128$ resolution using four methods on the Austria model at 5~GHz. Columns from left to right correspond to ground truth, CSI-TV, MRCSI, DeepCSI-Interp., and DeepCSI-Query, respectively.}
    \label{fig_contrast_superres}
\end{figure}
The super-resolution advantage extends to the reconstructed permittivity. Given the queried contrast source at $128 \times 128$, the bilinear interpolation of the low-resolution permittivity serves as an initial estimate. This estimate is then refined by 50 iterations of TV-regularized optimization that exploits the high-frequency information in the queried contrast source. This refinement step takes approximately 1~second. As reported in Table~\ref{tab_contrast_superres} and Fig.~\ref{fig_contrast_superres}, DeepCSI-Query achieves 20\% lower RMSE than DeepCSI-Interp. Furthermore, DeepCSI-Query trained on the $32 \times 32$ grid outperforms both CSI-TV and MRCSI directly inverted on the $128 \times 128$ grid, while requiring $140\times$ less computation time. This demonstrates that the continuous representation decouples the inversion resolution from the training resolution, enabling high-fidelity super-resolution at a fraction of the computational cost.

\subsubsection{Initialization Stability Analysis}
 
Phaseless inversion is notoriously prone to local minima due to the loss of phase information. To evaluate the sensitivity of DeepCSI to initialization, 10~independent phaseless single-frequency inversions at 3~GHz with 5\% noise are performed per test case. Different random seeds control only the network weight initialization, while the noise realization and all other settings are kept fixed. DeepCSI is compared against the Neural-$E^{\mathrm{tot}}$ baseline.

\begin{figure}[t]
    \centering
    \includegraphics[width=\linewidth]{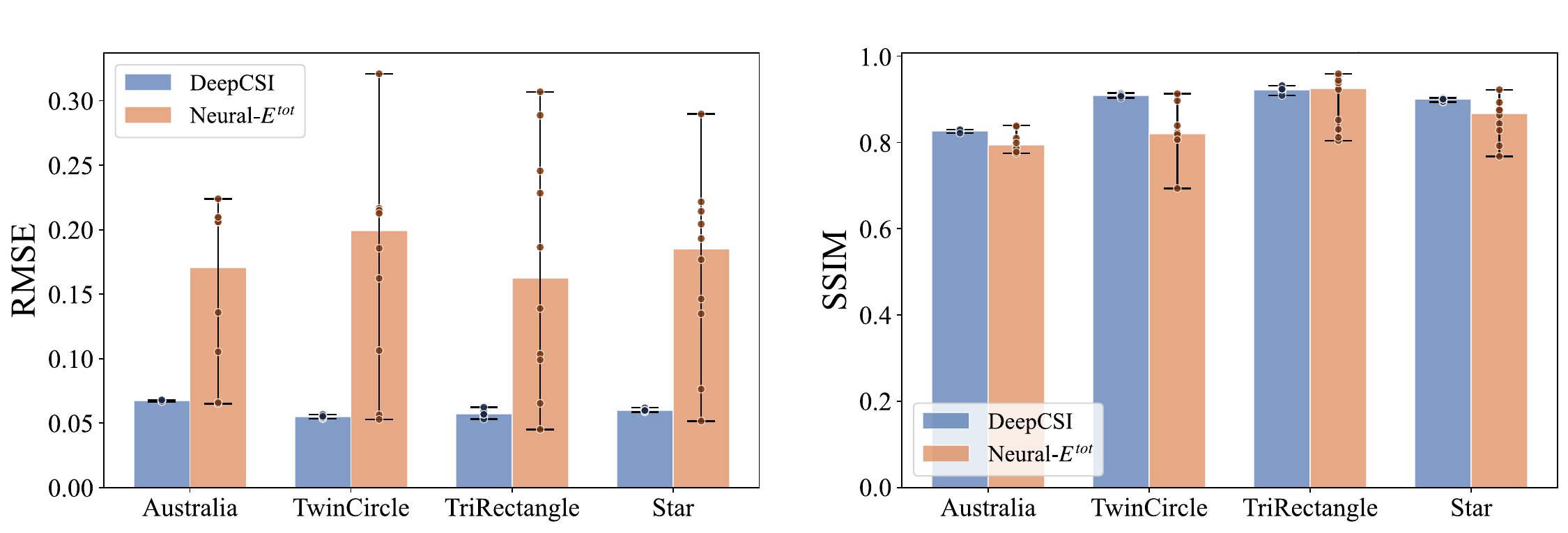}
    \caption{RMSE and SSIM of 10 independent runs with different random seeds under PD-SF inversion at 3~GHz with 5\% noise, for DeepCSI and Neural-$E^{\mathrm{tot}}$ across four representative cases.}
    \label{stability_boxplot}
\end{figure}

\begin{figure}[t]
    \centering
    \includegraphics[width=\linewidth]{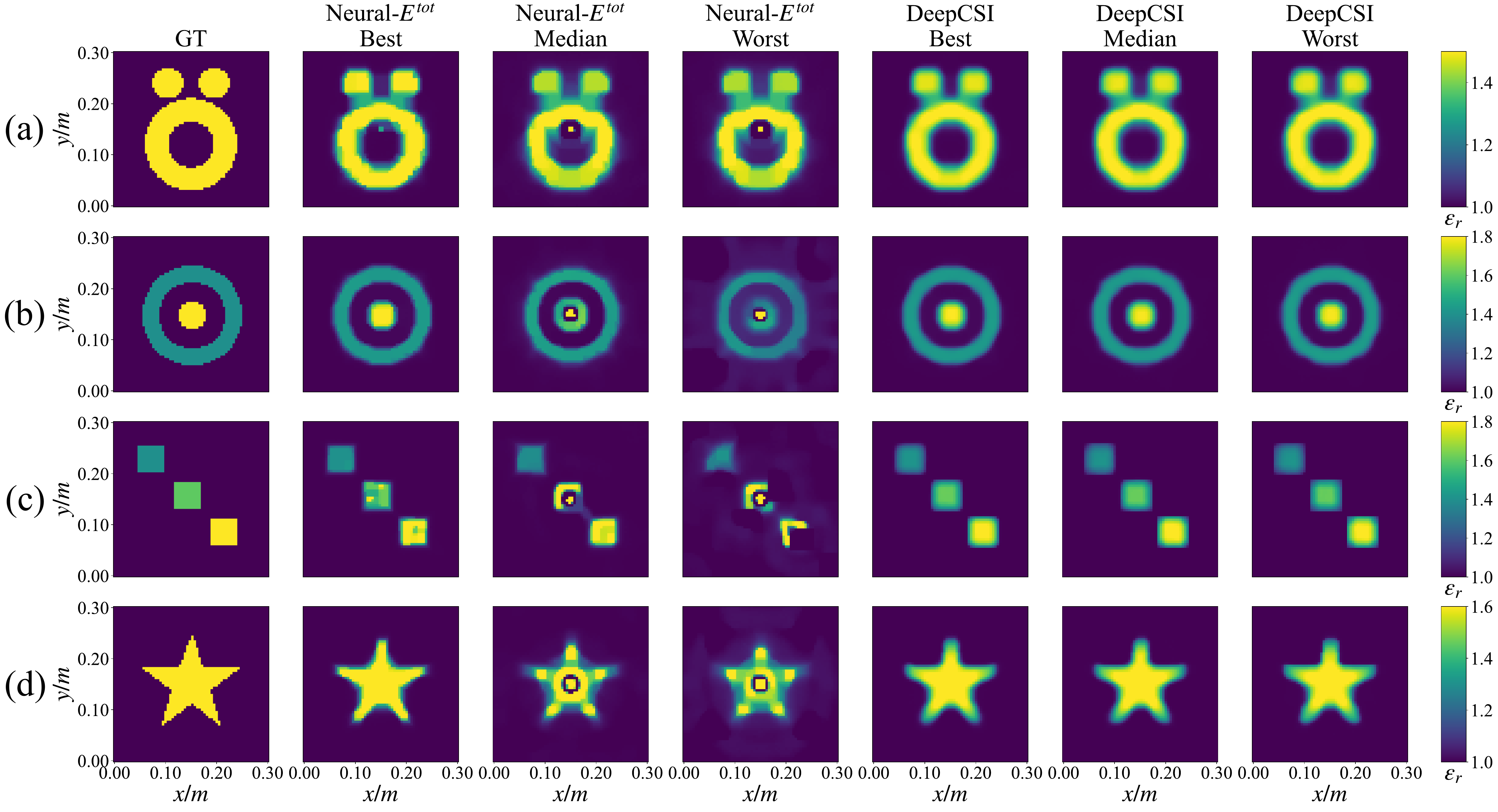}
    \caption{Best, median, and worst reconstruction results across 10 random seeds under PD-SF inversion at 3~GHz with 5\% noise for DeepCSI and Neural-$E^{\mathrm{tot}}$.}
    \label{stability_visual}
\end{figure}

As shown in Fig.~\ref{stability_boxplot}, DeepCSI achieves nearly identical RMSE and SSIM values across all 10 seeds for each test case, indicating near-deterministic convergence regardless of initialization. In contrast, the Neural-$E^{\mathrm{tot}}$ baseline exhibits large variations across seeds, with individual runs ranging from near-optimal to severely degraded. This instability reflects the non-uniqueness of the Eq.\eqref{eq:omega_chi} decomposition, which creates multiple equivalent local minima under phaseless conditions. The best, median, and worst reconstructions across 10~seeds are further visualized in Fig.~\ref{stability_visual}, where DeepCSI's outputs are visually indistinguishable, while the baseline's worst cases exhibit severe artifacts. These results confirm that DeepCSI can be reliably deployed as a single-run method without multiple restarts or initialization tuning.

\subsection{Experimental Data Inversion}
In this section, the proposed DeepCSI method is further validated using the Fresnel experimental data. The experimental configurations are detailed in \cite{geffrin2005free}. All TM-polarized measured data of the $FoamDielExt$, $FoamDielInt$, and $FoamTwinDiel$ models are selected for inversion. Both full-data and phaseless-data inversions are conducted based on the Fresnel measurements to evaluate the performance of DeepCSI in real-world scenarios.

The DOI is defined as 0.17 $\text{m}$ $\times$ 0.17 $\text{m}$ and discretized into 64 $\times$ 64 grids, consistent with the configuration used in the synthetic inversions. The single-frequency inversions are conducted at 3 GHz, while the multi-frequency inversions use data at 3, 4, and 5 GHz.

\begin{figure}[t]
	\centering
	\includegraphics[width=0.45\textwidth]{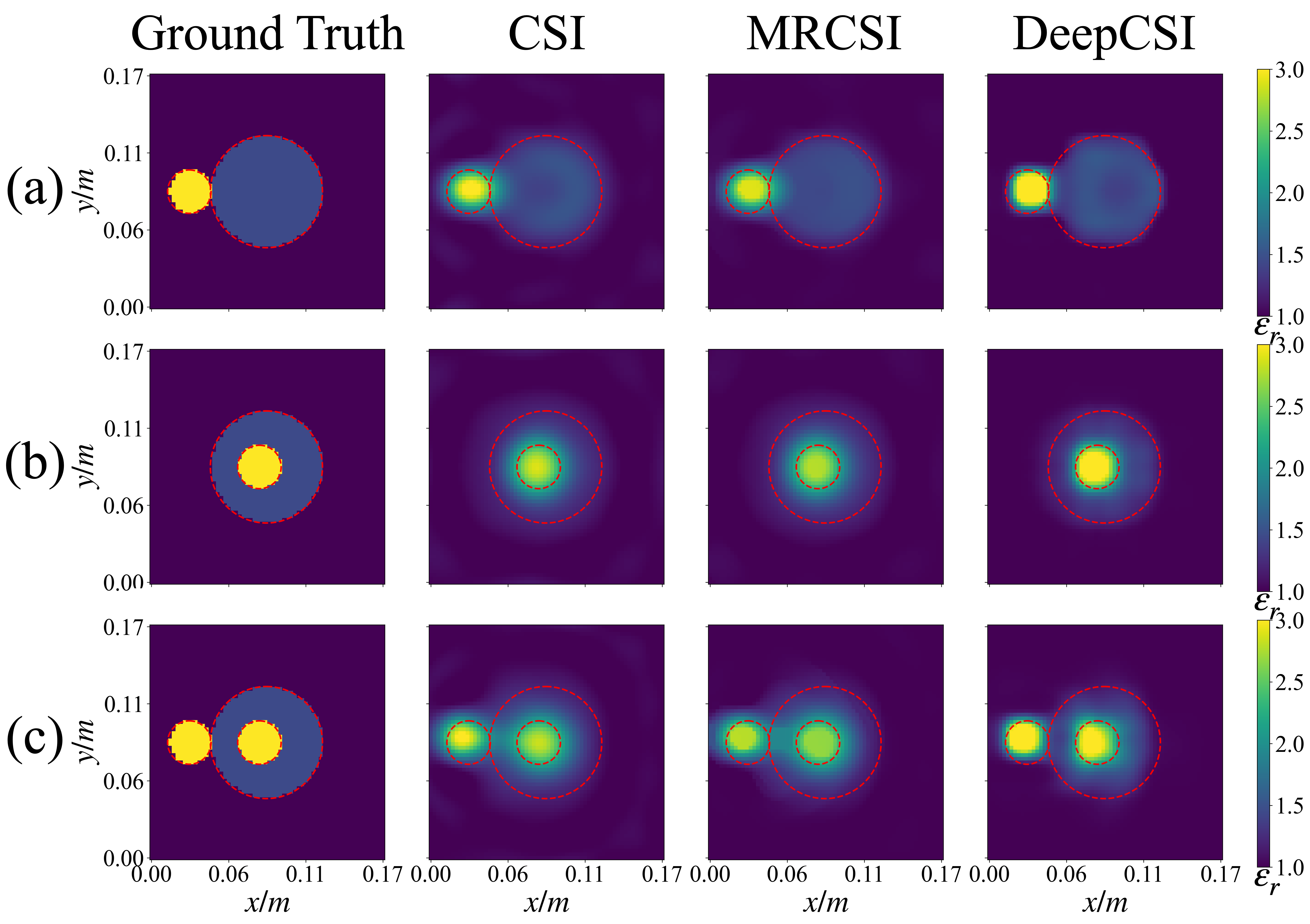}
	\caption{FD-SF inversion results for CSI, MRCSI, and DeepCSI on Fresnel data at 3 GHz. The first column represents the ground truth, while columns 2 to 4 represent the inversion results of CSI, MRCSI, and DeepCSI. (a)$\sim$(c) represents FoamDielExt, FoamDielInt, and FoamTwinDiel, respectively. The red dashed line indicates the real position and size of targets.}
	\label{figs_fresnelsingle}
\end{figure}

\begin{figure}[t]
	\centering
	\includegraphics[width=0.4\textwidth]{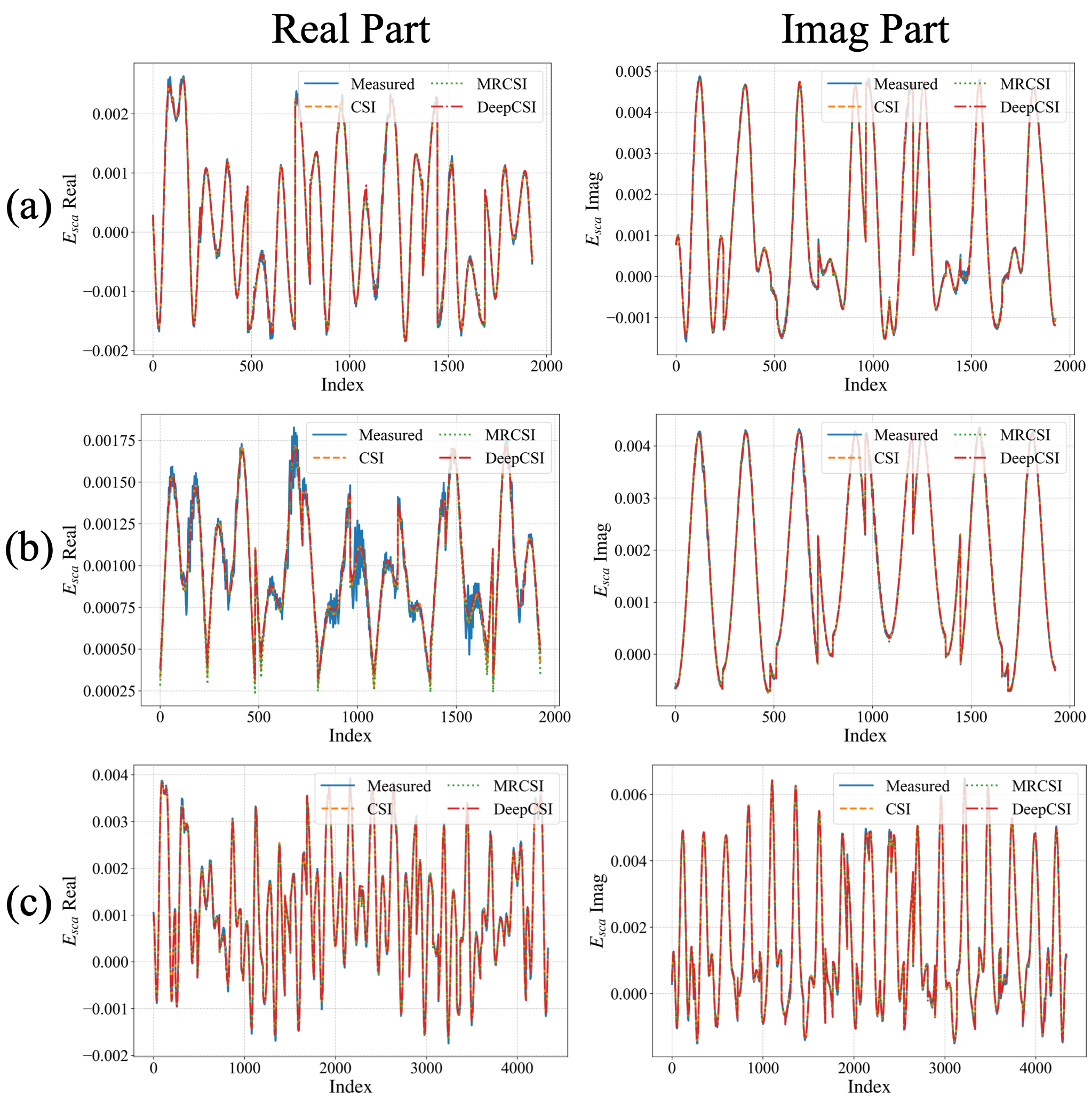}
	\caption{Comparison of measured and inverted scattered field data  using CSI, MRCSI, and DeepCSI on Fresnel full data at 3 GHz. The first and second columns represent the real and imag parts of scattered field, respectively. (a)$\sim$(c) represents FoamDielExt, FoamDielInt and FoamTwinDiel. The measured and inverted scattered field data are marked with different line styles.}
	\label{figs_fresnelsinglesca}
\end{figure}

\begin{table}[t]
	\centering
	\caption{FD-SF inversion performance of CSI, MRCSI, and DeepCSI on Fresnel data at 3 GHz}
	\label{tab_fresnelsingle}
	\begin{tabular}{c|>{\centering\arraybackslash}p{1.5cm}>{\centering\arraybackslash}p{1.2cm}>{\centering\arraybackslash}p{1.2cm}}
		\hline
		Case No.           & Method  & RMSE$\downarrow$   & SSIM$\uparrow$   \\ \hline
		\multirow{3}{*}{(a)} & CSI     & 0.1874 & 0.5985 \\
		& MRCSI   & 0.1870 & 0.7364 \\
		& DeepCSI & \textbf{0.1359} & \textbf{0.8435} \\ \hline
		\multirow{3}{*}{(b)} & CSI     & 0.1314 & 0.5511 \\
		& MRCSI   & 0.1302 & 0.6756 \\
		& DeepCSI & \textbf{0.1078} & \textbf{0.8204} \\ \hline
		\multirow{3}{*}{(c)} & CSI     & 0.1973 & 0.5498 \\
		& MRCSI   & 0.1997 & 0.6937 \\
		& DeepCSI & \textbf{0.1730} & \textbf{0.7770} \\ \hline
	\end{tabular}
\end{table}

\begin{figure}[t]
	\centering
	\includegraphics[width=0.4\textwidth]{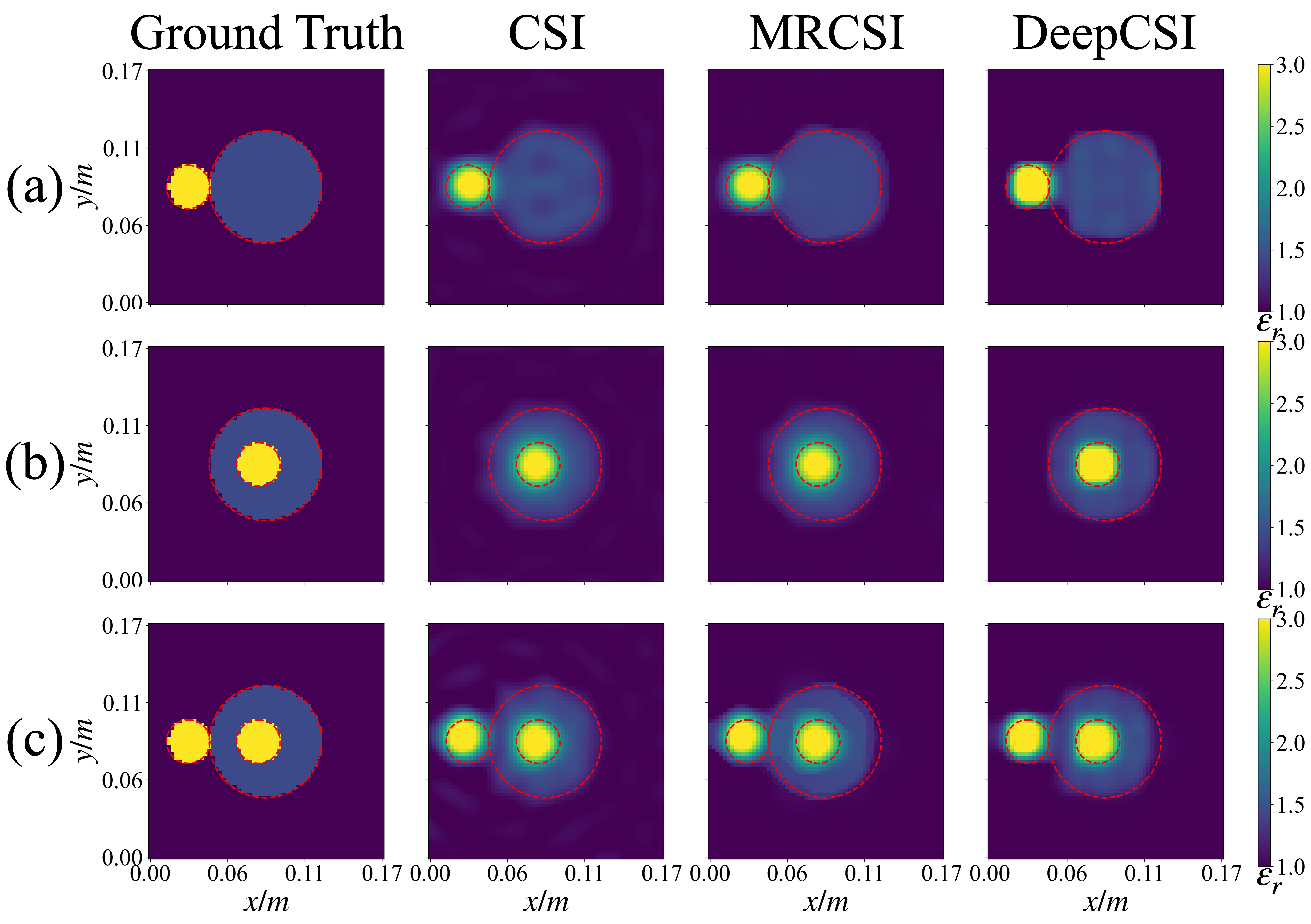}
	\caption{FD-MF inversion results for CSI, MRCSI, and DeepCSI on Fresnel data at 3, 4, and 5 GHz. The first column represents the ground truth, while columns 2 to 4 represent the inversion results of CSI, MRCSI, and DeepCSI. (a)$\sim$(c) represents FoamDielExt, FoamDielInt, and FoamTwinDiel, respectively. The red dashed line indicates the real position and size of targets.}
	\label{figs_fresnelmulti}
\end{figure}

\begin{table}[!t]
	\centering
	\caption{FD-MF inversion performance of CSI, MRCSI, and DeepCSI on Fresnel data at 3, 4, and 5 GHz}
	\label{tab_fresnelmulti}
	\begin{tabular}{c|>{\centering\arraybackslash}p{1.5cm}>{\centering\arraybackslash}p{1.2cm}>{\centering\arraybackslash}p{1.2cm}}
		\hline
		Case No.           & Method  & RMSE$\downarrow$   & SSIM$\uparrow$   \\ \hline
		\multirow{3}{*}{(a)} & CSI     & 0.1889 & 0.6232 \\
		& MRCSI   & 0.1867 & 0.8049 \\
		& DeepCSI & \textbf{0.1161} & \textbf{0.8668} \\ \hline
		\multirow{3}{*}{(b)} & CSI     & 0.1162 & 0.7106 \\
		& MRCSI   & 0.1158 & 0.8140 \\
		& DeepCSI & \textbf{0.0920} & \textbf{0.8923} \\ \hline
		\multirow{3}{*}{(c)} & CSI     & 0.1768 & 0.5867 \\
		& MRCSI   & 0.1766 & 0.7674 \\
		& DeepCSI & \textbf{0.1584} & \textbf{0.8230} \\ \hline
	\end{tabular}
	\vspace{-0.5em}
\end{table}

\subsubsection{Full Data Inversion}
 The FD-SF inversion results of CSI, MRCSI and DeepCSI at 3 GHz are shown in Fig. \ref{figs_fresnelsingle}. Compared to CSI and MRCSI, DeepCSI achieves higher inversion accuracy. Especially for the small cylinders with high contrast, DeepCSI can reconstruct their edge positions and medium parameters more accurately. Furthermore, the inverted scattered field data for each case is computed and compared with the measured scattered field data, as shown in Fig. \ref{figs_fresnelsinglesca}. It is evident that the inverted data align well with the measured data, demonstrating that DeepCSI can effectively fit scattered field data, highlighting its reliability in SF inversion. The RMSE and SSIM of the results from these methods are summarized in Table \ref{tab_fresnelsingle}. DeepCSI consistently outperforms CSI and MRCSI in both metrics. In particular, averaged over the three Fresnel targets, DeepCSI improves SSIM by 16.0\% and reduces RMSE by 19.3\% compared with MRCSI. All methods exhibit slightly worse inversion accuracy in case (c), probably due to the lower quality of measured data at 3 GHz compared with others.

The FD-MF inversion results of CSI, MRCSI and DeepCSI using 3, 4, and 5 GHz data are shown in Fig. \ref{figs_fresnelmulti}, and the performance metrics are presented in Table \ref{tab_fresnelmulti}. It can be seen that the inversion accuracy of all three methods improves under MF conditions, with DeepCSI still outperforming CSI and MRCSI. As shown in Fig. \ref{figs_fresnelmulti}, the inversion results using DeepCSI exhibit clear target contours and accurate medium parameters. In case (b) and (c), the inversion results of the high contrast small cylinder from DeepCSI show more accurate sizes, positions, and permittivity, which are highly consistent with the ground truth. This further validates the effectiveness of DeepCSI in MF inversion. In particular, for Fresnel full-data inversion, DeepCSI improves SSIM by 12.1\% on average and reduces RMSE by 21.1\% on average compared with MRCSI.

\begin{figure}[t]
	\centering
	\includegraphics[width=0.4\textwidth]{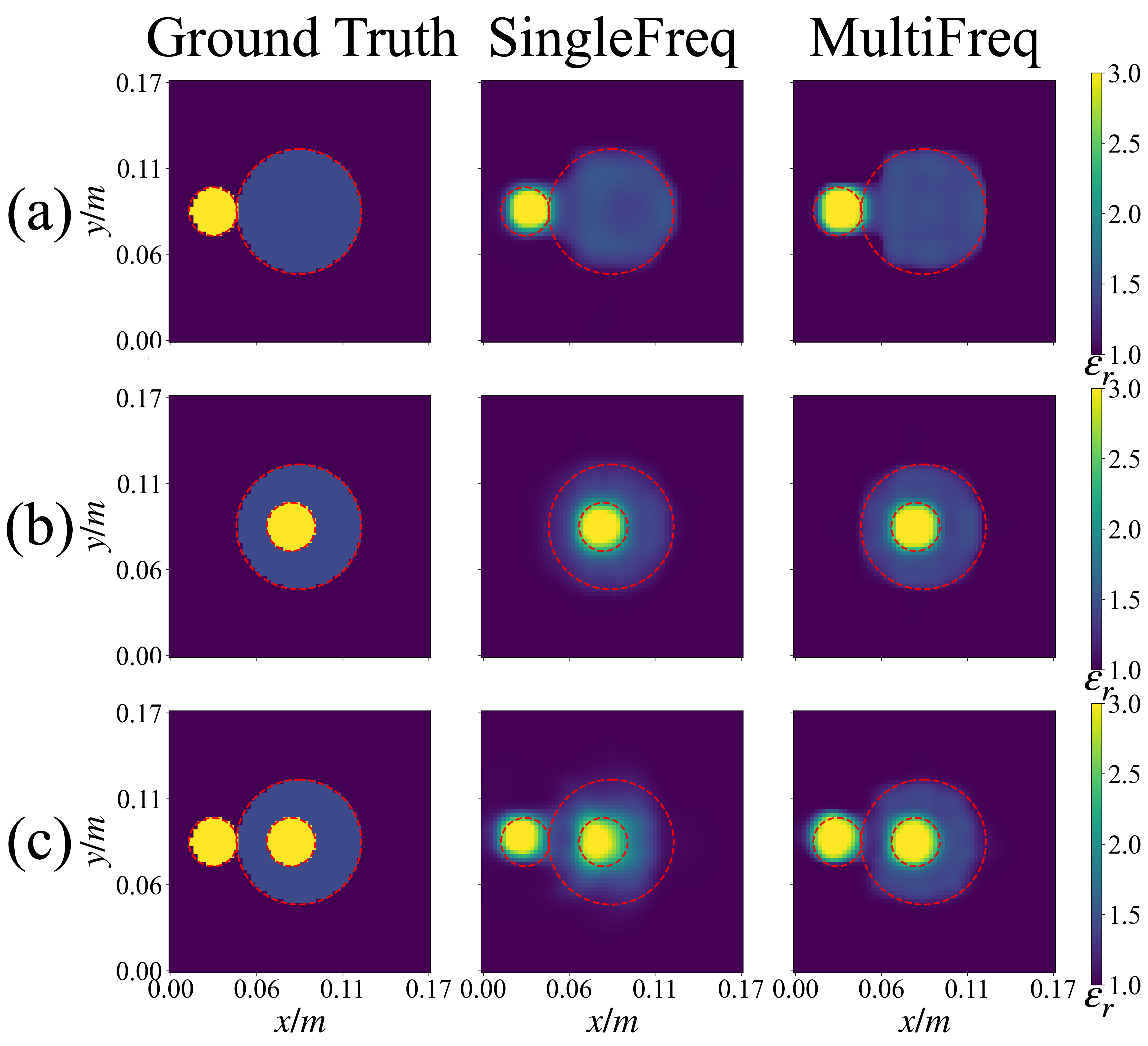}
	\caption{PD inversion results of DeepCSI on Fresnel data under SF and MF measurements. The first column represents the ground truth, while columns 2 to 3 represent the SF and MF inversion results. (a)$\sim$(c) represents FoamDielExt, FoamDielInt, and FoamTwinDiel, respectively. The red dashed line indicates the real position and size of targets.}
	\label{figs_fresnelpd}
\end{figure}

\subsubsection{Phaseless Data Inversion}
DeepCSI is further applied to the Fresnel phaseless data to validate its performance. Both single- and multi-frequency inversions are performed, and the results for the FoamDielExt, FoamDielInt, and FoamTwinDiel targets are presented in Fig.~\ref{figs_fresnelpd}, with quantitative metrics summarized in Table~\ref{tab_pd_fresnel}.

From the SF inversion at 3 GHz, DeepCSI accurately reconstructs the target geometry and medium parameter for FoamDielExt and FoamDielInt, while minor deviations appear in FoamTwinDiel. Using MF data at 3, 4, and 5 GHz yields clearer boundaries and more consistent dielectric profiles, confirming that DeepCSI maintains reliable reconstruction performance across frequencies.

\begin{figure}[t]
	\centering
	\includegraphics[width=0.5\textwidth]{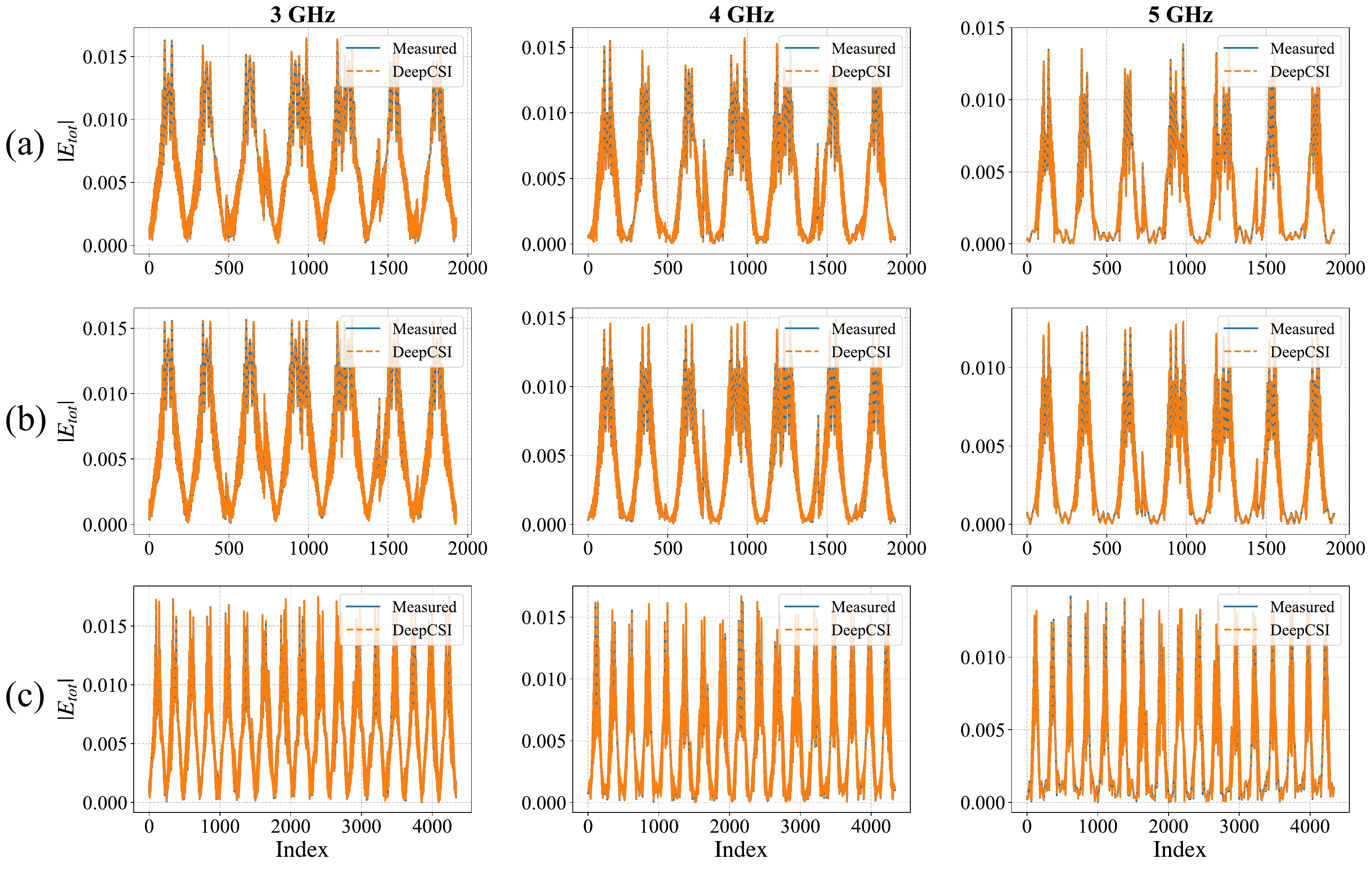}
	\caption{Comparison of measured and inverted total-field magnitudes using DeepCSI on Fresnel phaseless data for three models at 3, 4, and 5 GHz. (a)–(c) correspond to FoamDielExt, FoamDielInt, and FoamTwinDiel, respectively. The measured and inverted total-field magnitude data are distinguished by different line styles.}
	\label{figs_fresnelsinglesca_pd}
\end{figure}

Fig.~\ref{figs_fresnelsinglesca_pd} compares the measured and inverted total-field magnitudes of the three Fresnel targets, showing close agreement over all observation angles and frequencies. These results demonstrate that DeepCSI achieves accurate and robust phaseless inversion in both single- and multi-frequency experiments.

\begin{table}[t]
	\centering
	\caption{PD inversion performance of DeepCSI under SF and MF measurements on Fresnel data}
	\label{tab_pd_fresnel}
	\begin{tabular}{c|>{\centering\arraybackslash}p{1.8cm}>{\centering\arraybackslash}p{1.2cm}>{\centering\arraybackslash}p{1.2cm}}
		\hline
		Case No.           & Method  & RMSE$\downarrow$   & SSIM$\uparrow$   \\ \hline
		\multirow{2}{*}{(a)} & SF-DeepCSI     & 0.1045 & 0.8445 \\
		& MF-DeepCSI & \textbf{0.0904} & \textbf{0.8991} \\ \hline
		\multirow{2}{*}{(b)} & SF-DeepCSI     & 0.1365 & 0.8429 \\
		& MF-DeepCSI & \textbf{0.1178} & \textbf{0.8694} \\ \hline
		\multirow{2}{*}{(c)} & SF-DeepCSI     & 0.1738 & 0.7704 \\
		& MF-DeepCSI & \textbf{0.1523} & \textbf{0.8396} \\ \hline
	\end{tabular}
	\vspace{-0.5em}
\end{table}

\section{Conclusion}

In this work, we have presented DeepCSI, a physics-informed deep contrast source inversion framework for electromagnetic inverse scattering. By parameterizing the contrast source as a continuous coordinate-conditioned neural field with a lightweight ResMLP and integrating the volume integral equation into a differentiable computational graph, DeepCSI enables the joint optimization of the network parameters and the medium contrast within a unified framework. The same formulation accommodates both full and phaseless measurements by modifying only the data-consistency term.

Comprehensive experiments validate the proposed framework from multiple perspectives. A factorial ablation study shows that the neural contrast source representation is the primary factor in artifact suppression, with DeepCSI outperforming all pixel-based variants even without explicit regularization. Comparisons with a total-field parameterization baseline and a PINN-Helmholtz baseline further show that both the VIE-based formulation and the direct $\omega$-parameterization are essential for achieving high reconstruction quality. In addition, the continuous neural field enables super-resolution inference that decouples the inversion resolution from the reconstruction resolution. Under phaseless conditions, DeepCSI also exhibits near-deterministic convergence, in clear contrast to the much higher sensitivity observed with the total-field parameterization.


Despite these promising results, several directions remain for future work. First, electrically large and high-frequency inversions remain challenging, as the coordinate-based ResMLP may still suffer from spectral bias and slow convergence. Second, although the initialization stability analysis shows that DeepCSI is robust under the tested conditions, more complex scattering scenarios may still benefit from improved initialization strategies or multi-start schemes. Third, the current independent-subnet design for multi-frequency inversion could be extended to shared-weight architectures with frequency conditioning, potentially improving inter-frequency consistency while reducing the parameter count for many closely spaced frequencies.

\bibliographystyle{IEEEtran}
\bibliography{reference}


 




\vfill

\end{document}